\documentstyle[12pt]{article}
\input epsf

\def\SECTION#1{\begin{center}{\large\bf #1}\end{center}}

\def\rulerheight{0.5pt}

\def\U1{$U(1)$}
\def\SU5{$SU(5)$}
\def\SO10{$SO(10)$}
\def\422{$SU(4)\otimes SU(2)_L \otimes SU(2)_R$}

\def\diag.{\hbox{diag.}}

\def\refeqn#1{(\ref{#1})}
\def\M_U{\hbox{$M_U$}\ }
\def\M_P{\hbox{$M_P$}\ }

\def\tanb{\hbox{$\tan \beta$}}

\def\MSSM+N{\hbox{MSSM+$\nu$}}
\def\etal{{\it et al.}}

\def\ibid{{\it ibid.}}

\def\TabChargesLMA{I}
\def\TabYukMXAprLMA{II}
\def\TabaAValuesLMA{III}
\def\TabYukMXnumLMA{IV}
\def\TabYukMXepsLMA{V}
\def\TabYukMznumLMA{VI}
\def\TabMllVmnsLMA{VII}
\def\TabNeutMassLMA{VIII}
\def\TabAnglesLMA{IX}
\def\TabmdmsLMA{X}

\def\TabClebsch{XI}

\def\TabYukMznumLOW{XII}
\def\TabMllVmnsLOW{XIII}
\def\TabNeutMassLOW{XIV}
\def\TabAnglesLOW{XV}

\def\TabChargesSMA{XVI}
\def\TabYukMXAprSMA{XVII}
\def\TabaAValuesSMA{XVIII}
\def\TabYukMXnumSMA{XIX}
\def\TabYukMznumSMA{XX}
\def\TabMllVmnsSMA{XXI}
\def\TabNeutMassSMA{XXII}
\def\TabAnglesSMA{XXIII}
\def\TabmdmsSMA{XXIV}

\def\TabU1Charges{I}

\begin{document}
\baselineskip 24pt
\newcommand{\sheptitle}
{NEUTRINO MASSES AND MIXING ANGLES \\
IN A REALISTIC STRING-INSPIRED MODEL}

\newcommand{\shepauthor}
{S. F. King
and 
M.Oliveira}

\newcommand{\shepaddress}
{Department of Physics and Astronomy, University of Southampton \\
        Southampton, SO17 1BJ, U.K}

\newcommand{\shepabstract}
{We analyse a supersymmetric string-inspired 
 model of all fermion masses and mixing angles
 based on the Pati-Salam $SU(4)\times SU(2)_L \times SU(2)_R$ 
 gauge group supplemented by a $U(1)_X$ flavour symmetry. 
 The model involves third family Yukawa unification and predicts
 the top mass and the ratio of the vacuum expectation values $\tan \beta$.
 The model also provides a successful description of
 the CKM matrix and predicts the masses of the down and strange quarks.
 However our main focus is on the neutrino masses
 and MNS mixing angles, and we show how the recent
 atmospheric neutrino mixing observed by Super-Kamiokande, and
 the MSW solution to the solar neutrino problem
 lead to important information about the flavour structure of the model
 near the string scale.
 We show how single right-handed neutrino dominance may be implemented
 by the use of ``Clebsch zeros'', 
 leading to the LMA MSW solution, corresponding to 
 bi-maximal mixing. The LOW MSW and SMA MSW solutions are also discussed.}

\begin{titlepage}
\begin{flushright}
hep-ph/0009287
\end{flushright}
\begin{center}
{\large{\bf \sheptitle}}
\\ \shepauthor \\ \mbox{} \\ {\it \shepaddress} \\ 
{\bf Abstract} \bigskip \end{center} \setcounter{page}{0}
\shepabstract
\begin{flushleft}
\today
\end{flushleft}
\end{titlepage}

\newpage

\SECTION {I. INTRODUCTION}

The problem of understanding the quark and lepton masses and
mixing angles represents one of the major unsolved questions
of the standard model. Recently additional information on
the fermion mass spectrum has come from the measurement of the
atmospheric neutrino
masses and mixing angles by Super-Kamiokande \cite{SKamiokandeColl}.
The most recent data disfavours mixing involving a sterile
neutrino, and finds a good fit for $\nu_{\mu} \rightarrow \nu_{\tau}$
mixing with $\sin^22\theta_{23}>0.88$ and a mass square splitting
$\Delta m^2_{23}$ in the $1.5-5\times 10^{-3} {\rm\ eV}^2$
range at 90\% CL \cite{HSobel}. 
Super-Kamiokande has also provided additional support for
solar neutrino mixing. The most recent Super-Kamiokande
data does not show a significant
day-night asymmetry and shows an energy independent
neutrino spectrum, 
thus it also disfavours the sterile neutrino mixing
hypothesis, the just-so vacuum oscillation hypothesis, and
the small mixing angle (SMA) MSW \cite{MSWMechanism}
solution \cite{YSuzuki}.
The preferred solution at the present time seems to be the
large mixing angle (LMA) MSW solution, although a similar
solution with a low mass splitting (the LOW) solution is also
possible. A typical point in the LMA MSW region is 
$\sin^22\theta_{12}\approx 0.75$, and $\Delta m^2_{12}\approx 2.5\times
10^{-5} {\rm\ eV}^2$ \cite{BaKrSm}.

If one accepts the recent data as evidence for neutrino masses and
mixing angles, then the obvious question is how these can be
accommodated in the standard model, or one of its supersymmetric 
extensions. The simplest possibility to account for the smallness
of the neutrino masses is the see-saw mechanism \cite{seesaw}
in which one introduces right-handed neutrinos which acquire 
very large Majorana masses at a super-heavy mass scale.
When one integrates out the right-handed neutrinos the ``normal sized''
Dirac Yukawa couplings, which connect the left-handed to the right-handed
neutrinos, are transformed into very small couplings
which generate very light effective left-handed physical Majorana neutrino masses.
Given the see-saw mechanism, it is natural to expect that the
spectrum of the neutrino masses will be hierarchical, since the 
Dirac Yukawa couplings in the charged fermion sector are observed
to be hierarchical, and if they are related to the Dirac neutrino
Yukawa couplings then they should also be hierarchical,
leading to hierarchical light Majorana masses.
\footnote{However this is not guaranteed due to the unknown
structure of the heavy Majorana matrix, and for example an inverted
neutrino mass hierarchy could result although this 
relies on some non-hierarchical couplings in the Dirac 
Yukawa matrix \cite{KiSi2}.}

Having assumed the see-saw mechanism and a hierarchical 
neutrino mass spectrum, the next question is how such large
(almost maximal) lepton mixing angles such as $\theta_{23}$ could
emerge? There are several possibilities that have been suggested
in the literature. One possibility is that it happens
as a result of the off-diagonal 23 entries in the
left-handed Majorana matrix being large, 
and the determinant of the
23 sub-matrix being accidentally small, leading to a neutrino
mass hierarchy with large neutrino mixing angles \cite{ElLeLoNa}.
Another possibility is that the neutrino mixing angles start
out small at some high energy scale, then get magnified 
by renormalization group (RG) running down to low energies
\cite{BaLePa:MTanimoto}. A third possibility is that the 
off-diagonal elements of the left-handed neutrino Majorana matrix
are large, but the 23 sub-determinant of the matrix is small
for a physical reason, as would be the case if a single right-handed
neutrino were providing the dominant contribution to the
23 sub-matrix \cite{KingSRND,SFKing1,SFKing2}. 
We shall refer to these three approaches
as the accidental, the magnification and the single right-handed
neutrino dominance (SRHND) mechanisms, respectively.
As we shall see, in the model under consideration, only the
SRHND mechanism provides a successful description of the 
atmospheric neutrino data, and the results in this paper
will rely on this mechanism.

A promising approach to understanding the fermion mass spectrum is within
the framework of supersymmetric (SUSY) unified theories.
Within the framework of such theories the quark and lepton masses
and mixing angles become related to each other, and it begins
to be possible to understand the spectrum. The simplest
grand unified theory (GUT) is $SU(5)$ but this theory in its
minimal version does not contain any right-handed neutrinos.
Nevertheless three right-handed neutrinos may be added, and in this theory
it is possible to have a large 23 element 
\footnote{We use Left-Right (LR) convention for Yukawa matrices
in this paper.} on the Dirac neutrino
Yukawa matrix without introducing a large 23 element into any of the
charged fermion Yukawa matrices. The problem of maintaining
a 23 neutrino mass hierarchy in these models may be solved for example by
assuming SRHND \cite{AlFe2}. Another possibility within the
framework of $SU(5)$ is to maintain all the off-diagonal
elements to be small, but require the 22 and 32 elements of
the Dirac neutrino Yukawa matrix to be equal and the
second right-handed neutrino to be dominant, in which case
SRHND again leads to a large 23 neutrino mixing angle with 
hierarchical neutrino masses \cite{AlFeMa}.
However the drawback of $SU(5)$ is that it does not predict
any right-handed neutrinos, which must be added as an afterthought.

From the point of view of neutrino masses, the most natural
GUTs are those like $SO(10)$ that naturally predict right-handed
neutrinos. However within the framework of $SO(10)$ the
quark masses and mixing angles are related to the lepton 
masses and mixing angles, and the existence
of large neutrino mixing angles is not expected in the minimal versions
of the theory in which the Higgs doublets are in one (or two) ${\bf 10}'s$
(ten dimensional representations of $SO(10)$) 
and each matter family is in a ${\bf 16}$.
Nevertheless various possibilities
have been proposed in $SO(10)$ in order to account for 
the large neutrino mixing angles. Within the framework of
minimal $SO(10)$ with third family Yukawa unification, 
it has been suggested that if two operators with different
Clebsch coefficients contribute with similar strength then,
with suitable choice of phases,
in the case of the lepton Yukawa matrices one may
have large numerical 23 elements, which add up to give a
large lepton mixing angle, while for the quarks the 23 elements can be
small due to approximate cancellation of the two contributing
operators \cite{BaPaWi}. This is an example of the accidental mechanism
mentioned above, where in addition one requires the quark mixing
angles to be small by accident, although it remains to be seen if
the LMA MSW solution could be understood in this framework.
Moving away from minimal $SO(10)$,
one may invoke a non-minimal Higgs sector in which one Higgs
doublet arises from a ${\bf 10}$ and one from a ${\bf 16}$, and in this 
framework it is possible to understand atmospheric neutrino
mixing \cite{AlBa}. 
Alternatively, one may invoke a non-minimal matter
sector in which parts of a quark and lepton family arise from
a ${\bf 16}$ and other parts from a ${\bf 10}$, and in these models 
one may account for atmospheric and solar neutrinos
via an inverted mass hierarchy mechanism \cite{ShTa}.

In the present paper we shall discuss 
neutrino masses and mixing angles in a particular string-inspired
{\em minimal} model based on the Pati-Salam 
$SU(4)\times SU(2)_L \times SU(2)_R$ (422) group \cite{PaSa}.
As in $SO(10)$ the presence of the gauged $SU(2)_R$ predicts the
existence of three right-handed neutrinos.
However, unlike $SO(10)$, there is no Higgs doublet-triplet
splitting problem since in the minimal model
both Higgs doublets are contained in a $(1,2,2)$ representation.
Moreover, since the left-handed quarks and leptons are in the
$(4,2,1)$ and the right-handed quarks and leptons in the $(4,1,2)$ 
representations,
the model also leads to third family
Yukawa unification as in minimal $SO(10)$.
Although the Pati-Salam gauge group is not unified at the
field theory level, it readily emerges from string constructions
either in the perturbative fermionic constructions \cite{AnLe},
or in the more recent type I string constructions \cite{ShTy},
unlike $SO(10)$ which typically requires large Higgs representations
which do not arise from the simplest string constructions.
The question of fermion masses and mixing angles in the
string-inspired Pati-Salam model has already been discussed
for the case of charged fermions \cite{SFking2,AlKi3},
and later for the case of neutrinos \cite{AlKi4}.
For the neutrino study \cite{AlKi4} it was assumed that the 
heavy Majorana neutrino mass matrix was proportional to the
unit matrix, and only small neutrino mixing angles were considered.
Later on a $U(1)_X$ family symmetry was added to the model,
in order to understand the horizontal hierarchies, although in this
case the neutrino spectrum was not analysed at all \cite{AlKiLeLo1}.

The purpose of the present paper is to discuss neutrino masses
and mixing angles in the string-inspired Pati-Salam model
supplemented by a $U(1)_X$ flavour symmetry. 
The model involves third family Yukawa unification and predicts
the top mass and the ratio of the vacuum expectation values $\tan \beta$,
as we recently discussed in Ref.~\cite{KiOl2}. It is already known that
the model can provide a successful description of
the CKM matrix and predicts the down and strange
quark masses, although our present
analysis differs from that presented previously \cite{AlKiLeLo1}
partly due to the recent refinements in third family Yukawa unification 
\cite{KiOl2}, but mainly as a result of the recent Super-Kamiokande data
which has important implications for the flavour structure of the model.
In fact our main focus here is on the neutrino masses
and mixing angles which were not previously discussed at all in this
framework. We assume a minimal version of the model, and avoid the use 
of the accidental cancellation mechanism, which in any case has
difficulties in accounting for bi-maximal neutrino mixing.
We also show that the mixing angle magnification mechanism
can only provide limited increases in the mixing angles, due
to the fact that the  unified third family Yukawa coupling
is only approximately equal to 0.7 \cite{KiOl2} and is therefore too
small to have a dramatic effect. Instead, we rely on the
SRHND mechanism, and we show how this mechanism
may be implemented in the 422 model by appropriate use of 
operators with ``Clebsch zeros''
resulting in a natural explanation for atmospheric neutrinos
via a hierarchical mass spectrum. We specifically
focus on the LMA MSW solution in the text,
with the LOW and SMA MSW solutions relegated to Appendices.

The layout of the remainder of the paper is as follows.
In section II we briefly review the see-saw mechanism in the
Minimal Supersymmetric Standard Model (MSSM) \cite{MSSM}
with right-handed neutrinos. In section III we review
some useful analytic results for SRHND, for the case of an
approximately diagonal right-handed Majorana mass matrix.
In section IV we introduce the string-inspired Pati-Salam model,
and in section V we introduce an Abelian anomalous gauge
$U(1)_X$ family symmetry into the model, and show how
horizontal Yukawa hierarchies may be generated.
In section VI we describe our operator approach to fermion
masses, including the heavy Majorana neutrino masses.
Section VII contains the main results of the paper.
In this section we show how a particular choice of 
$U(1)_X$ family charges, and operators with certain Clebsch
coefficients can lead to a successful description of 
quark and lepton masses and mixing angles, and in particular
describe atmospheric and solar neutrinos via SRHND.
Although the neutrino masses and mixing angles
correspond to the usual LMA MSW solution, in Appendix~C
we show how a modification of the heavy Majorana mass matrix
can lead to a large mixing angle MSW solution with a LOW mass splitting.
In Appendix~D we present a different choice of $U(1)_X$
charges and operators which can lead to the SMA MSW solution.

\SECTION{II. THE MSSM WITH \\ RIGHT-HANDED NEUTRINOS}

The superpotential of the MSSM with right-handed neutrinos is given by :
\begin{eqnarray}
{\cal W} &=& {\cal W}_{MSSM}+{\cal W}_{\nu^c} \label{SRNDSuperPot} \\
{\cal W}_{MSSM} &=&
  q_A (\lambda_u)_{AB} u^c_B h_u 
- q_A (\lambda_d)_{AB} d^c_B h_d 
- l_A (\lambda_e)_{AB} e^c_B h_d 
+ \mu h_u h_d \\
{\cal W}_{\nu^c} &=&
  l_A (\lambda_\nu)_{AB} \nu^c_B h_u
+ {\textstyle {1 \over 2}} \nu^c_A (M_{RR})_{AB} \nu^c_B 
\end{eqnarray}
where $A,B=1,..,3$ are family indices,
$u^c$, $d^c$, $e^c$ and $\nu^c$ are the right-handed $SU(2)_L$ singlet 
superfields,
$q=(u,d)$ and $l=(\nu,e)$ are the $SU(2)_L$ quark and lepton doublets, and 
$h_u$ ($h_d$) is the up (down) Higgs boson doublet. 
The Dirac neutrino coupling and the heavy Majorana mass for 
the right-handed neutrinos are denoted by $\lambda_\nu$ and $M_{RR}$ 
respectively.
When the neutral components of the two MSSM Higgs bosons $h^0_{u,d}$ acquire 
their vacuum expectation values (VEV)s $v_{2,1}$
($\tan\beta=v_2/v_1 \sim 40-50$) the superpotential 
in Eq.~\refeqn{SRNDSuperPot} 
generates the following sum of mass terms :
\begin{eqnarray}
{\cal L}_{U,D,E} &=&-U (\lambda_u v_2) {U^c}
                    -D (\lambda_d v_1) {D^c}
                    -E (\lambda_e v_1) {E^c} + {\rm h.c.} \label{SRNDLagUDE} \\
{\cal L}_{N}     &=&-N (\lambda_\nu v_2) {N^c}
            -{\textstyle {1\over 2}} {N^c} M_{RR} {N^c} + {\rm h.c.}
\label{SRNDLagN}
\end{eqnarray}
where the upper case letters now denote the fermionic
components of the superfields in ${\cal W}$,
for example $u$ contains $(U,\tilde u)\equiv (U_L,\tilde u_L)$
and $u^c$ contains $(U^c, \tilde u^c)\equiv(U^*_R,\tilde u^*_R)$.
The Yukawa matrices in Eq.~\refeqn{SRNDLagUDE} can be diagonalized by bi-unitary
transformations $S$ and $T$ defined by :
\begin{equation}
T^{u*} \lambda_u S^{uT} = \lambda'_u \qquad
T^{d*} \lambda_d S^{dT} = \lambda'_d \qquad
T^{e*} \lambda_e S^{eT} = \lambda'_e
\end{equation}
Thus the physical (primed) states $U'_{R,L}$ are related to the 
gauge eigenstates $U_{R,L}$ by $U'_R = S^u U_R$ and $U'_L = T^u U_L$,
{\it etc}..
In this model, the left-handed neutrino masses are generated 
via see-saw mechanism \cite{seesaw} by the terms in Eq.~\refeqn{SRNDLagN} which 
can be re-arranged into a two-by-two block matrix in the following way :
\begin{equation}
{\cal L}_N = - {\textstyle {1\over 2}}
                ( N \, N^c )
                \left( \matrix{ 0      & m_{LR} \cr
                               m^T_{LR} &   M_{RR} }\right)
                \left( \matrix{ N \cr N^c} \right) + {\rm h.c.}
\end{equation}
where $m_{LR} = \lambda_\nu v_2$. 
Thus, after the heavy $N^c$ fields are integrated out,
the light left-handed neutrinos $N$ effectively acquire a small
mass given by :
\begin{equation}
m_{LL} = m_{LR} M^{-1}_{RR}\, m_{LR}^T
\label{SRNDSeeSaw}
\end{equation}
Finally, the diagonalization of $m_{LL}$ :
\begin{equation}
T^{N*} m_{LL} T^{N\dagger}_{L} = 
{\rm diag}(m_{\nu_1},m_{\nu_2},m_{\nu_3})
\end{equation}
allows the determination of the masses of the physical neutrinos
$m_{\nu_A}$ and enables the physical neutrino states 
$N' = (\nu_1,\nu_2,\nu_3)$ to be related to the neutrino gauge fields 
$N = (\nu_e, \nu_\mu, \nu_\tau)$ by $N' = T^N N$.

Taking into account the above conventions, we now proceed to give
expressions for the Cabbibo-Kobayashi-Maskawa (CKM) matrix 
\cite{CKM} ($V^{CKM}$)
and the corresponding lepton analogue, the Maki-Nakawaga-Sakata (MNS) 
matrix \cite{MaNaSa} ($V^{MNS}$).
Their definitions derive from the charged current interactions :
\footnote{The four component fermion fields $\Psi$ are given by 
$\Psi_F   = (F , -i\sigma^2 F^{c*})$ for $F=U,D,E$ 
and
$\Psi_{N} = (N, -i\sigma^2 N^{*})$
for the neutrinos.}
\begin{equation}
- {g \over \sqrt{2}} W^+_\mu \bar\Psi_U \gamma^\mu P_L \Psi_D \to
- {g \over \sqrt{2}} W^+_\mu \Psi_{U'} \gamma^\mu P_L V^{CKM} \Psi_{D'}
\end{equation}
\begin{equation}
- {g \over \sqrt{2}} W^-_\mu \bar\Psi_E \gamma^\mu P_L \Psi_{N} \to
- {g \over \sqrt{2}} W^-_\mu \bar\Psi_{E'} \gamma^\mu P_L V^{MNS} \Psi_{N'}
\end{equation}
that imply :
\begin{equation}
V^{CKM} = T^u T^{d\dagger} \qquad
V^{MNS} = T^e T^{N\dagger}
\label{SRNDCKMMNS}
\end{equation}
In what follows we will assume that the matrices in
Eq.~\refeqn{SRNDCKMMNS} are real.
\footnote{We shall not address the
question of CP violation in this paper.} 
Thus, we will write $V^{MNS}$ in terms of three rotation matrices :
\begin{equation}
V^{MNS} = R_{23} R_{13} R_{12}
\label{MNSRot}
\end{equation} 
given by :
\begin{equation}
R_{23} = \left( \matrix{ 1 & 0 & 0 \cr
                         0 & \phantom{-}c_{23} & s_{23} \cr
                         0 & -s_{23} & c_{23} } \right) \quad
R_{13} = \left( \matrix{ \phantom{-}c_{13} & 0 & s_{13} \cr
                             0  & 1 & 0 \cr
                        -s_{13} & 0 & c_{13} } \right) \quad
R_{12} = \left( \matrix{ \phantom{-}c_{12} & s_{12} & 0 \cr
                        -s_{12} & c_{12} & 0 \cr
                             0  &     0  & 1 } \right)
\label{RotMtr}
\end{equation}
where $s_{AB} = \sin\theta_{AB}$, $c_{AB} = \cos\theta_{AB}$ refer to the 
lepton mixing angles between the $A$ and $B$ generation. 
Using Eq.~\refeqn{RotMtr} into Eq.~\refeqn{MNSRot} gives :
\begin{equation}
V^{MNS} =  \left( \matrix{
                  c_{12} c_{13} &
                  s_{12} c_{13} &
                  s_{13} \cr
                  -s_{12} c_{23}-c_{12} s_{23} s_{13} & 
                  \phantom{-} c_{12} c_{23}-s_{12} s_{23} s_{13} &
                   s_{23} c_{13} \cr
                  \phantom{-} s_{12} s_{23}-c_{12} c_{23} s_{13} &
                  -c_{12} s_{23}-s_{12} c_{23} s_{13} &
                   c_{23} c_{13} } \right)
\label{VMNS}
\end{equation}
It is also practical to have expressions for the $\theta_{AB}$ angles in 
terms of the $V^{MNS}$ entries. Inverting Eq.~\refeqn{VMNS} we find
that :
\footnote{
$V_{e2} = V^{MNS}_{12}$, 
$V_{e3} = V^{MNS}_{13}$ and 
$V_{\mu 3} = V^{MNS}_{23}$.}
\begin{equation}
\sin\theta_{13} = V_{e3} \qquad
\sin\theta_{23} = {V_{\mu 3} \over \sqrt{1-V_{e3}^2}} \qquad
\sin\theta_{12} = {V_{e2} \over \sqrt{1-V_{e3}^2}}
\end{equation}

Finally we note that while the above expressions were derived in the
context of three neutrino species, the analysis of the experimental
results assumed only two, thus a direct comparison of mixing angles 
is not exactly valid.

\SECTION{III. SINGLE RIGHT-HANDED NEUTRINO DOMINANCE}

Third family single right-handed neutrino dominance (SRHND) 
\cite{KingSRND,SFKing1,SFKing2}
is a mechanism that can explain the large
atmospheric ($\theta_{23}$) and the solar LMA MSW ($\theta_{12}$) 
neutrino mixing angles and a small $\theta_{13}$.
SRHND relies on the possibility that the neutrino mass matrix
($m_{LL}$) is dominated by the contributions coming solely from a
single right-handed neutrino (for example $\nu^c_\tau$.)
In this scheme a maximal $\theta_{23}$ angle arises when the
tau right-handed neutrino $\nu^c_\tau$ couples to the left-handed muon
$\nu_\mu$ and tau neutrino $\nu_\tau$ with equal strength.
Similarly, if $\nu^c_\mu$ couples to $\nu_e$ and to $\nu_\mu$
with comparable strength then $\theta_{12}$ is large.
The role of the (sub-dominant) muon neutrino
is also important since it provides small perturbations to the
$m_{LL}$ matrix (which otherwise has one heavy and two massless eigenstates),
thus leading to a neutrino mass splitting 
$\Delta m^2_{12}=|m^2_{\nu_2}-m^2_{\nu_1}|$ compatible
with experiment.
In this section we summarize the theory behind SRHND and review
the analytic results presented in Ref.~\cite{SFKing2} 
for the case of the
diagonal dominated right-handed neutrino mass matrix $M_{RR}$.

The see-saw formula for the left-handed neutrino matrix in 
Eq.~\refeqn{SRNDSeeSaw} depends explicitly on $M_{RR}$. 
Although $M_{RR}$ might have a non-trivial structure we 
find instructive to start our analysis by considering the very simple
case of $M_{RR}$ given by :
\begin{equation}
M^{-1}_{RR} \sim {\rm diag}(M^{-1}_{\nu_1},M^{-1}_{\nu_2},M^{-1}_{\nu_3}) \sim
                 {\rm diag}(0,0,M^{-1}_{\nu_3}) 
\label{SRNDmRR}
\end{equation}
which effectively corresponds to taking $M_{\nu_1},M_{\nu_2} \gg M_{\nu_3}$.
Replacing Eq.~\refeqn{SRNDmRR} into Eq.~\refeqn{SRNDSeeSaw} we find that :
\footnote{In this section we will use the following simplified
notation $m_{LR} = \lambda_\nu v_2 \equiv \lambda v_2 \sim \lambda$.}
\begin{equation}
m_{LL} = 
v_2^2 \lambda_\nu M^{-1}_{RR} \lambda^T_\nu
\sim
\lambda M^{-1}_{RR} \lambda^T
\sim
M^{-1}_{\nu_3} 
\left(\matrix{
\lambda_{13}^2 & \lambda_{13} \lambda_{23} & \lambda_{13} \lambda_{33} \cr
\lambda_{13} \lambda_{23} & \lambda_{22}^2 & \lambda_{23} \lambda_{33} \cr
\lambda_{13} \lambda_{33} & \lambda_{23} \lambda_{33} & \lambda_{33}^2 \cr}
\right)
\label{SRNDmLLMTR1}
\end{equation}
The $m_{LL}$ matrix above is easily diagonalized by the matrices 
$R_{23}$, $R_{13}$, $R_{12}$
\footnote{Note that $R_{12}$ for $m_{LL}$ in Eq.~\refeqn{SRNDmLLMTR1}
is undetermined.}
in Eq.~\refeqn{RotMtr} with
rotation angles given by :
\begin{equation}
\matrix{
\displaystyle
s_{23} = {\lambda_{23} \over A} &
\displaystyle
c_{23} = {\lambda_{33} \over A} & &
A^2 = \lambda_{33}^2+\lambda_{23}^2  \hfill \cr
& & \qquad & \cr
\displaystyle
s_{13} = {\lambda_{13} \over B} &
\displaystyle
c_{13} = {           A \over B} & &
\displaystyle
B^2 = \lambda_{33}^2+\lambda_{23}^2+\lambda_{13}^2
}
\label{SRNDRotAngles}
\end{equation}
that successively act on $m_{LL}$ as follows :
\begin{equation}
m_{LL}^{\prime\prime\prime} = 
R_{12}^\dagger R_{13}^\dagger R_{23}^\dagger m_{LL} R_{23} R_{13} R_{12} =
{\rm diag}(m_{\nu_1},m_{\nu_2},m_{\nu_3})
\end{equation}
It is also convenient to define the following primed matrices :
\begin{equation}
m_{LL}^{\prime} = R_{23}^\dagger m_{LL} R_{23} \quad\quad
m_{LL}^{\prime\prime} = R_{13}^\dagger m_{LL}^{\prime} R_{13} \quad\quad
m_{LL}^{\prime\prime\prime} = R_{12}^\dagger m_{LL}^{\prime\prime} R_{12}
\label{SRNDmLLprime}
\end{equation}
which, for $m_{LL}$ as in Eq.~\refeqn{SRNDmLLMTR1}, are explicitly given by :
\begin{equation}
m_{LL}^\prime \sim M^{-1}_{\nu_3} \left(\matrix{
\lambda_{13}^2 & 0 & \lambda_{13} A \cr
0 & 1 & 0 \cr
\lambda_{13} A & 0 & A^2}\right)\qquad
m_{LL}^{\prime\prime} \equiv m_{LL}^{\prime\prime\prime} 
\sim M^{-1}_{\nu_3} \left(\matrix{
0 & 0 & 0 \cr 0 & 0 & 0 \cr 0 & 0 & B^2}\right)\quad
\label{SRNDmLLprimeMTR}
\end{equation}

We can see from Eq.~\refeqn{SRNDRotAngles} that if $\lambda_{23} =
\lambda_{33}$ then a maximal $\theta_{23}=45^0$ angle results.
Moreover, if $\lambda_{13} \ll \lambda_{23},\lambda_{33}$ 
then $\theta_{13}$ is small.
Although SRHND, in the limiting case of Eq.~\refeqn{SRNDmRR}, 
is successful in predicting a maximal atmospheric neutrino angle, 
it fails to account for a viable neutrino spectrum.
Indeed, from Eq.~\refeqn{SRNDmLLprimeMTR}, we see that the two
lightest neutrinos are massless $m_{\nu_1} = m_{\nu_2} = 0$.
Moreover the solar neutrino angle $\theta_{12}$ is undetermined.
These two problems can be solved by allowing the right-handed muon neutrino
$\nu^c_\mu$ to play a sub-dominant/perturbative role in the structure 
of $m_{LL}$ in Eq.~\refeqn{SRNDmLLMTR1}.

We now turn to the more realist model in which $M_{RR}$ can be
approximated by :
\footnote{Note that although Eq.~\refeqn{SRNDmLLMTR2} still looks very
simple it can, in many cases, provide a good qualitative 
description of the physics involving the heaviest neutrinos. 
Indeed, if $M_{RR}$ is diagonal dominated and if $m_{RL}$ is highly
hierarchical then the limiting case of Eq.~\refeqn{SRNDmLLMTR2} applies.}
\begin{equation}
M^{-1}_{RR} \sim 
{\rm diag}(M^{-1}_{\nu_1},M^{-1}_{\nu_2},M^{-1}_{\nu_3}) \sim
{\rm diag}(0,M^{-1}_{\nu_2},M^{-1}_{\nu_3})
\label{SRNDmLLMTR2}
\end{equation}
Using Eq.~\refeqn{SRNDmLLMTR2} into Eq.~\refeqn{SRNDSeeSaw}
we find that :
\medskip
\begin{equation}
m_{LL} \sim 
M^{-1}_{\nu_3}
\left(\matrix{
\lambda_{13}^2 &
\lambda_{13}\lambda_{23} &
\lambda_{13}\lambda_{33} \cr
\lambda_{13}\lambda_{23} &
\lambda_{23}^2 &
\lambda_{23}\lambda_{33} \cr
\lambda_{13}\lambda_{33} &
\lambda_{23}\lambda_{33} &
\lambda_{33}^2
}\right)+
M^{-1}_{\nu_2}
\left(\matrix{
\lambda_{12}^2 & 
\lambda_{12}\lambda_{22} &
\lambda_{12}\lambda_{32} \cr
\lambda_{12}\lambda_{22} &
\lambda_{22}^2 &
\lambda_{22}\lambda_{32} \cr
\lambda_{12}\lambda_{32} &
\lambda_{22}\lambda_{32} &
\lambda_{32}^2
}\right)
\label{SRNDmLLM2M3}
\end{equation}

\noindent
Given that we assumed SRHND by the $\nu^c_\tau$ neutrino, it follows
that the contributions to the 23 block of $m_{LL}$ in Eq.~\refeqn{SRNDmLLM2M3}
arising from the terms proportional to $M^{-1}_{\nu_3}$ dominate over 
the ones proportional to $M^{-1}_{\nu_2}$.
\footnote{Note that this does not necessarily imply that 
$M^{-1}_{\nu_3}$ is larger than $M^{-1}_{\nu_2}$
since the Yukawa couplings must also be taken into account.}
Clearly, the rotations 
$R_{12}$, $R_{13}$ parameterised by the angles in Eq.~\refeqn{SRNDRotAngles}
diagonalize $m_{LL}$ in Eq.~\refeqn{SRNDmLLM2M3} 
up to terms of order ${\cal O}(M^{-1}_{\nu_2})$.
Thus the new primed matrices $m^{\prime}_{LL}$ and 
$m^{\prime\prime}_{LL}$ are given by :
\begin{eqnarray}
m_{LL}^{\prime} &\sim& 
M^{-1}_{\nu_3}
\left(\matrix{
\lambda_{13}^2 &
0 &
\lambda_{13} A \cr
0 &
0 &
0 \cr
\lambda_{13} A &
0 &
A^2
}\right)+
M^{-1}_{\nu_2}
\left(\matrix{
\lambda_{12}^2 &
\lambda_{12} {C^2 \over A} &
\lambda_{12} {D^2 \over A} \cr
\lambda_{12} {C^2 \over A} &
 {C^4 \over A^2} &
 {C^2 D^2  \over A^2} \cr
\lambda_{12} {D^2 \over A} &
 {C^2 D^2 \over A^2} &
 {D^4 \over A^2}
}\right) \\
& & 
\nonumber \\
m_{LL}^{\prime\prime} &\sim&
M^{-1}_{\nu_3}
\left(\matrix{
0 & 0 & 0 \cr
0 & 0 & 0 \cr
0 & 0 & B^2
}\right)+
M^{-1}_{\nu_2}
\left(\matrix{
{E^6 \over A^2 B^2} &
{C^2 E^3 \over A^2 B} &
{F^2 E^3 \over A B^2} \cr
{C^2 E^3 \over A^2 B} &
{C^4 \over A^2} &
{C^2 F^2 \over AB} \cr
{F^2 E^3 \over A B^2} &
{C^2 F^2 \over AB} &
{F^4 \over B^2}
}\right)
\label{SRNDmLLpp}
\end{eqnarray}
where
\begin{eqnarray}
C^2 &=& \lambda_{22}\lambda_{33}-\lambda_{32}\lambda_{23} \\
D^2 &=& \lambda_{33}\lambda_{32}+\lambda_{22}\lambda_{23} \\
E^3 &=& \lambda_{12} (\lambda_{33}^2+\lambda_{23}^2) - 
        \lambda_{13} (\lambda_{33}\lambda_{32}+\lambda_{22}\lambda_{23}) \\
F^2 &=& \lambda_{33}\lambda_{32}+\lambda_{22}\lambda_{23}+\lambda_{12}\lambda_{13}
\end{eqnarray}
The diagonalization of the 12 block of $m^{\prime\prime}_{LL}$ 
in Eq.~\refeqn{SRNDmLLpp} is achieved by a $R_{12}$ matrix parameterised by the
following $\theta_{12}$ rotation angle :
\begin{equation}
s_{12} ={E^3 \over \sqrt{E^6+B^2 C^4}} \qquad
c_{12} ={B C^2 \over \sqrt{E^6+B^2 C^4}}
\end{equation}
Thus we find :
\begin{equation}
m^{\prime\prime\prime}_{LL} \sim
M^{-1}_{\nu_3}
\left(\matrix{
0 & 0 & 0 \cr
0 & 0 & 0 \cr
0 & 0 & B^2
}\right)+
M^{-1}_{\nu_2}
\left(\matrix{
0 & 0 & 0 \cr
0 & {E^6+B^2 C^4 \over A^2 B^2} & {F^2 \sqrt{E^6+B^2 C^4} \over A B^2} \cr
0 &  {F^2 \sqrt{E^6+B^2 C^4} \over A B^2} & {F^4 \over B^2}
\label{SRNDmLLppp}
}\right)
\end{equation}
It is interesting to note that the $R_{12}$ rotation has
not only diagonalized the 12 block of $m^{\prime\prime}_{LL}$ but also
put zeros in the 13,31 entries of $m^{\prime\prime\prime}_{LL}$. 
The reason is because $m^{\prime\prime}_{LL}$ displays a special
structure. Indeed, their elements obey :
\begin{equation}
t_{12} = 
{s_{12} \over c_{12}} =
{(m^{\prime\prime}_{LL})_{12} \over (m^{\prime\prime}_{LL})_{22}} =
{(m^{\prime\prime}_{LL})_{11} \over (m^{\prime\prime}_{LL})_{12}} =
{(m^{\prime\prime}_{LL})_{13} \over (m^{\prime\prime}_{LL})_{23}} =
{E^3 \over B C^2}
\end{equation}
Explicitly $t_{12} = \tan\theta_{12}$ is given by :
\begin{equation}
t_{12} = 
{\lambda_{12}(\lambda_{33}^2+\lambda_{23}^2)-
 \lambda_{13}(\lambda_{33}\lambda_{32}+\lambda_{22}\lambda_{23}) \over
(\lambda_{22}\lambda_{33}-\lambda_{32}\lambda_{23})
 \sqrt{\lambda_{33}^2+\lambda_{23}^2+\lambda_{13}^2}}
\sim {\lambda_{12} \over \lambda_{22}}
\label{SRNDt12}
\end{equation}
From Eq.~\refeqn{SRNDt12} we see that, although $t_{12}$ generally
depends on the second and third family neutrino Yukawa couplings, 
if $\lambda_{33}$ is much bigger than the other Yukawa couplings
then $t_{12} \sim \lambda_{12} / \lambda_{22}$.
This means that the $\theta_{12}$ angle is set not by the dominant
neutrino couplings, but by the sub-dominant $\nu^c_\mu$ neutrino 
couplings to the $\nu_e$ and $\nu_\mu$ neutrinos.
Thus while a large atmospheric neutrino mixing angle $\theta_{23}$ can
be achieved by requiring $\lambda_{23}\sim\lambda_{33}$, 
a large MSW solar neutrino angle $\theta_{12}$ results from 
$\lambda_{12}\sim\lambda_{22}$. 
Moreover, Eq.~\refeqn{SRNDRotAngles} and Eq.~\refeqn{SRNDt12} show that
bi-maximal $\theta_{23}$, $\theta_{12}$ mixing can be achieved with a small 
$\theta_{13}$ angle as long as 
$\lambda_{13} \ll \lambda_{23},\lambda_{33}$.
The neutrino mass spectrum can be read from 
$m^{\prime\prime\prime}_{LL}$ in Eq.~\refeqn{SRNDmLLppp}.
We find a massless neutrino state $m_{\nu_1} = 0$, 
plus a light state with mass $m_{\nu_2} \sim \lambda^2_{22} / M_{\nu_2}$ 
and a heavy neutrino with mass $m_{\nu_3} \sim \lambda^2_{33} / M_{\nu_3}$.

\SECTION{IV. THE PATI-SALAM MODEL}

Here we briefly summarize the parts of the Pati-Salam model
\cite{PaSa} that are relevant for our analysis.
For a more complete discussion see Ref.~\cite{AnLe}.
The SM fermions, together with the right-handed neutrinos, are conveniently
accommodated in the following $F=(4,2,1)$ and $F^c=(\bar 4,1,\bar 2)$
representations :
\begin{equation}
\qquad
F_A   = \left(\matrix{
            u & u & u & \nu \cr
            d & d & d & e \cr}
            \right)_A
\qquad
F^c_B = \left(\matrix{
            d^c & d^c & d^c & e^c \cr
            u^c & u^c & u^c & \nu^c \cr}
            \right)_B
\label{FcFields}
\end{equation}
The MSSM Higgs bosons fields are contained in $h=(1,\bar 2,2)$ :
\begin{equation}
h = \left(\matrix{h_d^- & h_u^0 \cr
                  h_d^0 & h_u^+ \cr} \right)
\label{SMHiggs}
\end{equation}
whereas the heavy Higgs bosons $\bar H=(\bar 4,1,\bar 2)$ and $H=(4,1,2)$
are denoted by :
\begin{equation}
\bar H = \left(\matrix{ \bar H_d & \bar H_d & \bar H_d & \bar H_e \cr
                        \bar H_u & \bar H_u & \bar H_u & \bar H_\nu
                        \cr}\right)
\qquad
H  = \left(\matrix{ H_d & H_d & H_d & H_e \cr
                    H_u & H_u & H_u & H_\nu  \cr}\right).
\label{RHiggs}
\end{equation}
In addition to the Higgs fields in Eqs.~\refeqn{SMHiggs},\refeqn{RHiggs}
the model also involves an $SU(4)$ sextet field $D=(6,1,1)=(D_3,D^c_3)$.

The superpotential of the minimal 422 model is :
\begin{eqnarray}
{\cal W} &=& F \lambda F^c h + \lambda_h S h h + \nonumber \\
         & & \lambda_S S (\bar H H - M^2_H) +
             \lambda_H H H D + \lambda_{\bar H} {\bar H} {\bar H} D +
             F^c \lambda' F^c \,{H H \over M_V}
\label{W3}
\end{eqnarray}
where $S$ denotes a gauge singlet superfield, the $\lambda$'s
are real dimensionless parameters and 
$M_H \sim M_X \sim 10^{16} {\rm\ GeV}$.
Additionally, $M_V > M_X$ denotes the mass of extra exotic matter
that has been integrated out from the model at high energy. 
As a result of the superpotential terms involving the singlet $S$,
the Higgs fields develop VEVs
$\langle H \> \rangle = \langle H_\nu \rangle \sim M_X$ and
$\langle \bar H \rangle = \langle \bar H_\nu \rangle \sim M_X$
which lead to the symmetry breaking~:
\begin{equation}
SU(4) \otimes SU(2)_L \otimes SU(2)_R \to
SU(3)_c \otimes SU(2)_L \otimes U(1)_Y.
\label{422:321}
\end{equation}
The singlet $S$ itself also naturally develops a small VEV of the order
of the SUSY breaking scale \cite{KiSh} so that the $\lambda_h S$ term in Eq.~\refeqn{W3}
gives an effective $\mu$ parameter of the correct order of magnitude.
Under Eq.~\refeqn{422:321} the Higgs field $h$ in Eq.~\refeqn{SMHiggs}
splits into the familiar MSSM doublets $h_u$ and $h_d$ whose neutral
components subsequently develop weak scale VEVs
$\langle h_u^0 \rangle = v_2$ and $\langle h_d^0 \rangle = v_1$ with
$\tanb = v_2/v_1$.
The neutrino fields $\nu^c$ acquires a large mass
$M_{RR} \sim \lambda' \langle HH \rangle / M_V$ through the
non-renormalizable term in ${\cal W}$ which, together
with the Dirac $\nu^c$ -- $\nu$ interaction
(proportional to $\lambda \langle h_u^0 \rangle$),
gives rise to a 2 $\times$ 2 matrix that generates, via see-saw
mechanism \cite{seesaw},
a suppressed mass for the left-handed neutrino states.
The $D$ field does not develop a VEV but the terms $HHD$ and $\bar H \bar H D$
combine the colour triplet parts of $H$, $\bar H$ and $D$ into acceptable
GUT scale mass terms \cite{AnLe}.

\SECTION{V. ABELIAN FLAVOUR SYMMETRY}

The pattern of fermion masses and mixing angles is one of the fundamental
problems is particle physics that has not yet been understood. 
The importance of this unsolved puzzle is demonstrated by 
the numerous works published in the literature over the past years 
(see Refs.~\cite{MatrixModels1}-\cite{BaDo} for a ``short'' list.)
In the standard model (SM) the quark/lepton masses and the CKM
matrix are input parameters fixed by laboratory experiments.
Surprisingly, however, their values, though unconstrained and {\it a priori}
arbitrary, do display a certain degree of organization. 
The fermion masses are highly hierarchical and the CKM matrix can be
described in terms of the small Wolfenstein expansion parameter 
$\lambda \sim |V_{12}| \sim 0.22$ \cite{LWolfenstein}.
These results suggest that a broken flavour symmetry might be playing
an important role in the setting of the structure of the Yukawa matrices.

In this work we will assume that the ``vertical'' gauge group is
supplemented by an additional $U(1)_X$ ``horizontal'' flavour symmetry
that constraints the nature of the couplings of quarks and leptons to
SM singlet fields $\theta$ and $\bar\theta$. 
The family symmetry, however, is broken at some high energy scale 
$M_\theta > M_X$ by the VEVs of the $\theta$, $\bar\theta$ fields which 
under the $U(1)_X$ group have charges $X_\theta = -1$ and
$X_{\bar\theta}= +1$.
As a consequence of the $U(1)_X$ symmetry breaking, the low energy
effective theory includes Dirac interactions between
the $F$ and $F^c$ fields of the following form :
\begin{equation}
F_A F^c_B h  \left( {\theta}
                    \over M_V \right)^{p_{AB}} \to
F_A F^c_B h \left( \langle\theta\rangle 
                   \over M_V \right)^{p_{AB}} \sim
F_A F^c_B h \epsilon^{p_{AB}}
\label{NonRenTheta}
\end{equation}
\begin{equation}
F_A F^c_B h  \left( {\bar\theta}
                    \over M_V \right)^{p_{AB}} \to
F_A F^c_B h \left( \langle\bar\theta\rangle 
                   \over M_V \right)^{p_{AB}} \sim
F_A F^c_B h \epsilon^{p_{AB}}
\label{NonRenThetaBar}
\end{equation}
where $p_{AB}$ is the modulos of the sum of the $U(1)_X$ charges
of the $F_A$, $F^c_B$ and $h$ fields, 
{\it i.e.} $p_{AB} = |X_{AB}| = |X_{F_A}+X_{F^c_B}+X_h|$. 
Thus Eq.~\refeqn{NonRenTheta} holds if $X_{AB} > 0$ 
whereas Eq.~\refeqn{NonRenThetaBar} holds if $X_{AB} < 0$.
The non-renormalizable terms in Eqs.~\refeqn{NonRenTheta},\refeqn{NonRenThetaBar}
might originate from interactions between the $F$ and $\theta$ fields 
with additional exotic vector matter with mass $M_V > M_X$ that 
lead to ``spaghetti'' diagrams as discussed in Ref.~\cite{AlKiSpaghetti}.
In summary, the equations above show that, in the context of a $U(1)_X$ symmetry,
the observed hierarchy in the fermion masses and mixing angles might
be the result of the flavour charges carried by the fields of the 422
model which act to suppress the Yukawa couplings by some $\epsilon$-power.

The introduction of the $U(1)_X$ symmetry provides a way to relate the
various flavour parameters of the model thus making it more predictive.
However, one should be careful. Generally the $U(1)_X$ group is
potentially dangerous since it can introduce, through triangle
diagrams, mixed anomalies with the SM gauge group.
\footnote{The cancellation of anomalies requires the vanishing of
the trace ${\rm Tr}(T^a \{T^b,T^c\}) = 0$ where $T^{a,b,c}$ are any of
the group generators which stand at the three gauge boson 
vertices of the triangle diagrams.}
In the last part of this section we review the constraints imposed on
$X$ charges of the fields of our model enforced by the
requirement of anomaly cancellation \cite{JaSh}.

The mixed anomalies that we shall consider are :
\footnote{We will not include the analysis of the $U(1)_X^3$ or of the
gravitational anomaly because they depend exclusively on SM singlet
fields.}
\begin{eqnarray}
\!\!\!\!\!\!\!\!
SU(3)^2 U(1)_X  & : &
A_3 = \sum_{A=1}^3 (2 X_{q_A}+ X_{u^c_A}+ X_{d^c_A}) \label{AnomA3} \\
\!\!\!\!\!\!\!\!
SU(2)^2 U(1)_X  & : &
A_2 = \sum_{A=1}^3 (3 X_{q_A}+ X_{l_A})+ X_{h_u}+ X_{h_d} \label{AnomA2} \\
\!\!\!\!\!\!\!\!
U(1)^2_Y U(1)_X & : &
A_1 = \sum_{A=1}^3 ({\textstyle {1\over 3}} X_{q_A}+
                      {\textstyle {8\over 3}} X_{u^c_A}+
                      {\textstyle {2\over 3}} X_{d^c_A}+
                       X_{l_A}+ 2 X_{e^c_A})+ X_{h_u}+ X_{h_d} \label{AnomA1} \\
\!\!\!\!\!\!\!\!
U(1)_Y U(1)_X^2  & : &
A'_1 = \sum_{A=1}^3 ( X_{q_A}^2- 2X_{u^c_A}^2+X_{d^c_A}^2
                      -X_{l_A}^2+ X_{e^c_A}^2) + X_{h_u}^2+ X_{h_d}^2 \label{AnomA1p}
\end{eqnarray}
For example, $A_3$ corresponds to the anomalous term generated by the
Feynman diagram that has two $SU(3)$ gluons and one $U(1)_X$ gauge
boson attached to the triangle vertices.
We note that the first three anomalies $A_3$, $A_2$ and $A_1$ are linear in
the trace of the charges, {\it i.e.} $X_f = \sum_{A=1}^3 X_{f_A}$, 
where $f$ is any of the $q,u^c,d^c,l,e^c$ fields, thus they constraint
only the family independent (FI) part of the $U(1)_X$ charges.
On the other hand, $A'_1$ is quadratic in the $X$ charges, 
thus it generally constraints the FI and family dependent (FD) part of 
the $U(1)_X$ charges.

In this paper we will assume that the cancellation of anomalies results
from the Green-Schwartz (GS) mechanism \cite{GrSc}.
This is possible if the $A_3$, $A_2$ and $A_1$ anomalies are in the
ratio $A_3:A_2:A_1=k_3:k_2:k_1$ where the $k_i$ are the Kac-Moody
levels of the $SU(3)$, $SU(2)$ and $U(1)_Y$ gauge groups that
determine the boundary conditions for the gauge couplings at the
string scale $g_3^2 k_3 : g_2^2 k_2 : g_1^2 k_1$.
Hence, using the canonical GUT normalization for the gauge couplings
(that successfully predicts $\sin^2(\theta_W)=3/8$ \cite{Ibanez}), 
anomalies can be
cancelled if we require that :
\begin{equation}
A_3=A_2={\textstyle {5 \over 3}}A_1
\label{AnomCanc}
\end{equation}
As a consequence of the two constraints implicit in Eq.~\refeqn{AnomCanc},
the set of solutions for the $X$ charges appearing 
in Eqs.~(\ref{AnomA3})-(\ref{AnomA1}) is given by \cite{JaSh} :
\begin{equation}
\matrix{
\displaystyle 
X_{e^c} = \sum_{A=1}^3 X_{e^c_A} = x \hfill &
\displaystyle 
X_{l^c} = \sum_{A=1}^3 X_{l_A}   = y \hfill &
\displaystyle 
X_{h_u} = -z \hfill \cr
\displaystyle 
X_{q\phantom{^c}}   = \sum_{A=1}^3 X_{q_A}   = x+u  \hfill &
\displaystyle 
X_{d^c} = \sum_{A=1}^3 X_{d^c_A} = y+v \hfill &
\displaystyle 
X_{h_d} = +z+(u+v) \hfill \cr
\displaystyle 
X_{u^c} = \sum_{A=1}^3 X_{u^c_A} = x+2u \hfill &
& }
\label{AnomSol}
\end{equation}
where $x,y,z,u,v$ are free parameters. 
However not all the solutions in Eqs.~\refeqn{AnomSol} are valid after 
$A'_1=0$ is enforced. In fact, as we said before, generally 
$A'_1$ constrains both the FI and FD charges of $U(1)_X$.
By this we mean that, if we conveniently write the charge of the $f_A$
field $X_{f_A}$ as a sum of a FI part $X_f$ plus a FD part $X'_{f_A}$,
{\it i.e.} $X_{f_A}= {1 \over 3} X_f+X'_{f_A}$, then $A'_1=0$ 
is a complicated equation on all $X_f$, $X'_{f_A}$ and $X_{h_u}$, $X_{h_d}$ charges.
However, it is easy to see that, if all the left-handed fields and 
if all the right-handed fields have the same FD charges,
{\it i.e.} $X'_{q_A}=X'_{l_A}$ and $X'_{u^c_A}=X'_{d^c_A}=X'_{e^c_A}$,
as is the case of the 422 model, then $A'_1=0$ is an equation on the
FI charges only :
\begin{equation}
A'_1 = {\textstyle {2 \over 3}}
(X^2_q- 2 X^2_{u^c}+ X^2_{d^c}- X^2_{l}+ X^2_{e^c}+ X^2_{h_u}-X^2_{h_d})=0
\end{equation}
Thus, a simple solution to all the anomaly constraints is given by  
Eq.~\refeqn{AnomSol} with $u=v=0$.
Finally, we must add that since the Pati-Salam model unifies all the 
left/right-handed quark and lepton fields in the $F$/$F^c$ multiplets, 
and the MSSM Higgs fields $h_u$, $h_d$ in the $h$ Higgs bi-doublet
we must also have $x=y$ and $z=0$. Thus, anomaly cancellation in the
422 model via GS mechanism is possible if the traces of the $U(1)_X$ 
charges of the $F$ and $F^c$ fields are equal,
{\it ie} $X_{F} = \sum_{A=1}^3 X_{F_A} \equiv \sum_{A=1}^3 X_{F^c_A} = X_{F^c}$. 

\SECTION{VI. OPERATOR APPROACH TO FERMION MASSES}

In the simplest formulation of the 422 model extended by 
a $U(1)_X$ horizontal symmetry all the Yukawa couplings 
originate from a single matrix.
The Abelian $U(1)_X$ symmetry introduced in the previous section
mainly serves one purpose, it establishes an hierarchy
between the flavour dependent couplings. Thus, it provides no
precise/predictive information about the relationships between the
different Yukawa coupling matrices.
As a result, all the SM fermions of a given family have
identical Yukawa couplings at the unification scale. 
Naturally, when the fermion masses are run from the $M_X$ to the $M_Z$ scale they lead 
to quark and lepton masses that are incompatible with the experimental data.

The idea of Yukawa unification, though unsuccessful in its most
simpler form, is not, however, a complete failure. As a matter of
fact, it turns out that third family Yukawa unification works rather
well. It is well known that the GUT boundary condition for
the Yukawa couplings :
\begin{equation}
\lambda_t(M_X)=\lambda_b(M_X)=\lambda_\tau(M_X)=\lambda_{\nu_\tau}(M_X)
\end{equation}
leads to a large pole top mass prediction $M_t \sim 175$ GeV and 
$\tan\beta\sim m_t/m_b$. On the other hand, the first and second
family fermion masses can be predicted if special 
relations between the ``vertical'' intra-generation Yukawa 
couplings at $M_X$ hold.
For example, the Georgi-Jarlskog (GJ) \cite{GeJa}
relation between the muon and strange Yukawa
couplings $\lambda_\mu \sim 3 \lambda_s$ successfully reproduces the low
energy experimental $m_s/m_\mu \sim 1$ mass ratio.
In the context of GUT theories the appearance of numerical factors
relating the couplings of the up-down-lepton Yukawa matrices might
originate from non-renormalizable operators involving the
interaction between the fermions and the heavy Higgs that break the
GUT symmetry \cite{FrNi,AnRaDiHaSt}.

In the Pati-Salam model, we will have in mind operators of the
following form \cite{AlKiLeLo1}~: 
\begin{equation}
F_A F^c_B h \left( H \bar H \over M_V^2 \right)^{n}
\left(\theta \over M_V \right)^{p_{AB}}
\quad{\rm\ and}\quad
F_A F^c_B h \left( H \bar H \over M_V^2 \right)^{n}
\left(\bar\theta \over M_V \right)^{p_{AB}}
\label{SRNDopRL}
\end{equation}
The idea is that when the $H$ and $\theta$ fields develop their
VEVs such operators reduce to effective Yukawa couplings with small coefficients. 
For example, if $F_2$, $F^c_2$ and $h$ carry a charge
$X_{F_2} = 0$, $X_{F^c_2} = 2$ and $X_h = 0$ under the $U(1)_X$
symmetry then Eq.~\refeqn{SRNDopRL} (with $n=1$) generates the following
terms :
\begin{equation}
(x_u u_2 u^c_2 h^0_u+x_d d_2 d^c_2 h^0_d+
 x_e e_2 e^c_2 h^0_d+x_\nu \nu_2 \nu^c_2 h^0_u)\delta\epsilon^2
\end{equation}
where $\delta=\langle H \rangle\langle\bar H \rangle/M_V^2$
and $\epsilon=\langle\theta\rangle / M_V$
are small dimensionless parameters,
$u_2$, $d_2$, $e_2$, $\nu_2$ are the charm, strange, muon, muon neutrino
superfields, 
and $x_f$ ($f=u,d,e,\nu$) are Clebsch factors that depend 
on the group theoretical contractions between the fields in 
Eq.~\refeqn{SRNDopRL} \cite{SFking2,AlKi3}.
In Table~\TabClebsch\ (Appendix A) 
we present a complete list of all $x_f$ values that result 
from $n=1$ operators in the 422 model \cite{AlKiLeLo1} normalized by :
\begin{equation}
x_u^2+x_d^2+x_e^2+x_\nu^2 = 4
\end{equation}
It is interesting to point out that different operators imply
zero Clebsches for different $x_f$'s. 
For example, CLASS-I operators are rather special since of all $x_f$'s 
only one is non-zero (and significantly large). 
The CLASS-II operators have $x_u=x_\nu=0$ while CLASS-III have $x_d=x_e=0$.
Additionally CLASS-IV operators have $x_u=x_d=0$ and CLASS-V have $x_e=x_\nu=0$. 
Finally CLASS-VI operators have all $x_f$'s different from zero. 
The variety of the operator Clebsches is to be welcome since, 
as we will see, they open the possibility of avoiding the disastrous fermion mass
predictions characteristic of the minimal 422 model with a unified
renormalizable interaction.

Finally we shall mention the origin of the heavy
Majorana neutrino mass matrix.
Generally $M_{RR}$ results from non-renormalizable operators of the form :
\begin{equation}
F^c_A F^c_B \left( H H \over M_V^2 \right)
\left( H \bar H \over M_V^2 \right)^n
\left(\theta \over M_V \right)^{q_{AB}}
\to
\nu^c_A \nu^c_B \delta^{n+1}\epsilon^{q_{AB}}
\label{SRNDopRR}
\end{equation}
where $q_{AB}=|X_{F^c_A}+X_{F^c_B}+\sigma|$ and $\sigma = 2 X_H$.
Three important differences distinguish Eq.~\refeqn{SRNDopRR} 
from Eq.~\refeqn{SRNDopRL}.
Firstly we note that while Eq.~\refeqn{SRNDopRL} allows for
renormalizable operators, $M_{RR}$ as given by Eq.~\refeqn{SRNDopRR} 
is always the result of non-renormalizable operators.
Secondly we note that the combination of the $HH$ fields in
Eq.~\refeqn{SRNDopRR} introduces an additional free parameter $\sigma$ 
that may be fixed at our convenience. Thirdly we observe that while
Eq.~\refeqn{SRNDopRL} is able to generate precise relationships between the
up-down-lepton Yukawa couplings (via Clebsch factors), 
Eq.~\refeqn{SRNDopRR} is an expression that constrains only the
hierarchy of $M_{RR}$ (via the $U(1)_X$ symmetry), 
as a result it is less predictive.

\SECTION{VII. NEUTRINO MASSES AND MIXING ANGLES}

In this section we show how the $U(1)_X$ horizontal family symmetry of
section V can be combined with the operator approach of section VI to
give predictions for the fermion masses and mixing angles in
the 422 model. 
In particular we are interested in the predictions for the neutrino
masses and mixing angles for the LMA MSW solution to the solar
neutrino problem.
The LOW and the SMA MSW solutions are discussed in
Appendices C and D.
We start by listing the quark and charged lepton 
experimental data used in our analysis :
\footnote{
The numbers inside the curly brackets indicate the 
experimental ranges according to Ref.~\cite{pdg}.}
\begin{eqnarray}
m_u({\rm 1\ GeV}) &=& 4.7  {\rm\ MeV} \qquad
\hfill
(1.35-6.75) {\rm\ MeV} \label{SRNDdatamu} \\
m_c(M_c)          &=& 1.21 {\rm\ GeV} \qquad
\hfill
(1.15-1.35) {\rm\ GeV} \\
M_t             & \sim & 175 {\rm\ GeV} \qquad
(170-180) {\rm\ GeV}
\end{eqnarray}
\begin{eqnarray}
m_d({\rm 1\ GeV}) & \sim & 6.0  {\rm\ MeV} \qquad
\hfill
(4-12) {\rm\ MeV} \\
m_s(M_s)          & \sim & 160     {\rm\ MeV} \qquad
\hfill
(100-230)  {\rm\ MeV} \\
m_b(M_b)          &=& 4.15    {\rm\ GeV} \qquad
\hfill
(4.0-4.4) {\rm\ GeV}
\end{eqnarray}
\begin{eqnarray}
M_e               &=& 0.511   {\rm\ MeV} \qquad 
m_e(M_e) = 0.496 {\rm\ MeV} \\
M_\mu             &=& 105.7   {\rm\ MeV} \qquad
m_\mu(M_\mu) = 104.6 {\rm\ MeV} \\
M_\tau            &=& 1777.0  {\rm\ MeV} \qquad
m_\tau(M_\tau) = 1772.8 {\rm\ MeV} \label{SRNDdatamtau}
\end{eqnarray}
where $m_u({\rm 1\ GeV})$, $m_d({\rm 1\ GeV})$ denote the running
masses of the up and down quarks at $Q=1{\rm\ GeV}$;
\footnote{All running masses are given in the $\overline{\rm MS}$ scheme.}
$m_c(M_c)$, $m_s(M_c)$, $m_b(M_b)$ the running masses of the charm,
strange and bottom quarks at their pole masses 
($M_c = 1.6 {\rm\ GeV}$, $M_b = 4.8 {\rm\ GeV}$);
$M_t$ the top pole mass; and $M_{e,\mu,\tau}$ 
($m_{e,\mu,\tau}$) the well known pole
(running) charged lepton masses.
We converted the above pole masses to running masses
using the expressions in Ref.~\cite{ArCaKeMiPiRaWr} with :
\begin{equation}
\alpha_s(M_Z) = 0.120 \qquad\qquad
\alpha_e^{-1}(M_Z) = 127.8
\end{equation}
Finally the CKM matrix at $Q=M_Z$ was fixed by :
\footnote{
The experimental ranges are
$|V_{12}| = 0.219 {\rm\ to\ } 0.226$, $|V_{23}|= 0.037 {\rm\ to\ } 0.043$ and 
$|V_{13}| = 0.002 {\rm\ to\ } 0.005$ \cite{pdg}.}
\begin{equation}
|V_{12}| = 0.2215 \qquad |V_{23}|= 0.040 \qquad |V_{13}| = 0.0035
\label{SRNDdataCKM}
\end{equation}
It is important to note that, in fact, not all the parameters above
were taken as an input. Indeed, $M_t \sim 175 {\rm\ GeV}$ is a
prediction that results from third family Yukawa unification 
$\lambda_t = \lambda_b = \lambda_\tau$ at the GUT scale.
Moreover, as we will see, our model is also able to predict
the masses of the down and charm quarks, thus their values listed
above should be taken merely as a guide and/or convenient initial estimates.

We now turn to the neutrino experimental data. 
The results from the Super-Kamiokande collaboration 
\cite{SKamiokandeColl,HSobel} indicate that the
atmospheric neutrino anomaly can be understood in terms of
$\nu_\mu\leftrightarrow\nu_\tau$ oscillations with :
\begin{equation}
\sin^2(2\theta_{23}) > 0.88 \qquad\qquad
1.5 \times 10^{-3}{\rm\ eV}^2 < \Delta m^2_{23} < 5\times 10^{-3} {\rm\ eV}^2
\label{SRNDSKdata}
\end{equation}
at 90 \% confidence level. On the other hand, the large mixing angle
(LMA) MSW solution to the solar neutrino deficit suggests that \cite{BaKrSm} :
\begin{equation}
\sin^2(2\theta_{12}) \sim 0.75 \qquad\qquad
\Delta m^2_{12} \sim 2.5 \times 10^{-5} {\rm\ eV}^2
\label{SRNDLMAMSWdata}
\end{equation}
Assuming that the neutrino spectrum is hierarchical, {\it i.e.} 
$\Delta m_{23}^2 = |m^2_{\nu_3}-m^2_{\nu_2}| \sim m^2_{\nu_3}$ and
$\Delta m_{12}^2 = |m^2_{\nu_2}-m^2_{\nu_1}| \sim m^2_{\nu_2}$ the values 
in Eqs.~\refeqn{SRNDSKdata},\refeqn{SRNDLMAMSWdata} give :
\begin{equation}
\sin(\theta_{23}) > 0.57 \quad m_{\nu_3}    \sim 0.05  {\rm\ eV}
\qquad\qquad
\sin(\theta_{12}) \sim 0.50 \quad m_{\nu_2} \sim 0.005 {\rm\ eV}
\label{SRNDneutrinoDATA}
\end{equation}
The latest results from the CHOOZ experiment also show 
that, over the interesting $\Delta m^2_{23}$ range suggested by the 
Super-Kamiokande data, $\sin^2(2\theta_{13})<0.10$ at 90\%~CL~\cite{CHOOZ}.
 
The experimental data in 
Eqs.~\refeqn{SRNDdatamu}-\refeqn{SRNDdataCKM}
constrains the parameters of our model at low energy. 
However, the GUT symmetry is broken at an energy $M_X \sim 10^{16} {\rm\ GeV}$. 
Thus, before we start our analysis we
should correct the fermion masses and mixing angles for the radiative
corrections that result from the running of the RGEs between the $Q=M_Z$ 
and $Q=M_X$ scales. The implementation of the RGEs, decoupling of SUSY
particles and boundary conditions is a complicated subject whose
detailed description is beyond the scope of this work. Here we will
only mention that we used 2-loop RGEs in the gauge and Yukawa
couplings and refer the interested reader to Ref.~\cite{KiOl2} 
where the issue of Yukawa unification in the 422 model is discussed. 
As a result of the RGEs running, subjected to third family Yukawa 
unification at $M_X$, the low energy input values for the
fermion masses in Eqs.~\refeqn{SRNDdatamu}-\refeqn{SRNDdatamtau} 
effectively constraint the eigenvalues of the Yukawa couplings at $Q=M_X$ to be :
\begin{equation}
\lambda_u(M_X) = 4.738\times 10^{-6} \qquad
\lambda_c(M_X) = 1.529\times 10^{-3} \qquad
\lambda_t(M_X) = 0.677
\label{SRNDdataMXUP}
\end{equation}
\begin{equation}
\lambda_d(M_X) \sim 3.208\times 10^{-4} \qquad
\lambda_s(M_X) \sim 9.612\times 10^{-3} \qquad
\lambda_b(M_X) = 0.677
\label{SRNDdataMXDOWN}
\end{equation}
\begin{equation}
\lambda_e(M_X)    = 1.490\times 10^{-4} \qquad
\lambda_\mu(M_X)  = 3.154\times 10^{-2} \qquad
\lambda_\tau(M_X) = 0.677
\label{SRNDdataMXLEP}
\end{equation}
and the CKM matrix at $Q=M_X$ to be :
\begin{equation}
|V_{12}(M_X)| = 0.2215 \qquad |V_{23}(M_X)|= 0.032 \qquad |V_{13}(M_X)| = 0.0028
\label{SRNDdataMXCKM}
\end{equation}
At this point it is convenient to re-write 
Eqs.~\refeqn{SRNDdataMXUP}-\refeqn{SRNDdataMXCKM} in terms of the
Wolfenstein \cite{LWolfenstein} 
expansion parameter $\lambda=0.22\sim|V_{12}|$. We find :
\begin{equation}
\lambda_u(M_X) = \lambda^{8.097} \qquad
\lambda_c(M_X) = \lambda^{4.282} \qquad
\lambda_t(M_X) = \lambda^{0.257}
\label{SRNDUPdata}
\end{equation}
\begin{equation}
\lambda_d(M_X) \sim \lambda^{5.313} \qquad
\lambda_s(M_X) \sim \lambda^{3.068} \qquad
\lambda_b(M_X) = \lambda^{0.257}
\end{equation}
\begin{equation}
\lambda_e(M_X)    = \lambda^{5.820} \qquad
\lambda_\mu(M_X)  = \lambda^{2.283} \qquad
\lambda_\tau(M_X) = \lambda^{0.257}
\end{equation}
\begin{equation}
|V_{12}(M_X)| = \lambda^{0.996} \qquad 
|V_{23}(M_X)| = \lambda^{2.273} \qquad 
|V_{13}(M_X)| = \lambda^{3.882}
\label{SRNDCKMdata}
\end{equation}
Equations \refeqn{SRNDUPdata}-\refeqn{SRNDCKMdata} 
neatly summarize the hierarchy of the quark
and charged lepton sectors at $M_X$ that we aim to reproduce/predict.

It is now time to specify the structure of the LMA model in more detail.
We start by indicating the nature of the (non-)renormalizable
operators responsible for the structure of the Dirac and neutrino 
Majorana matrices :
\begin{eqnarray}
\lambda_{AB} &:&
F_3 F^c_3 h +
F_A F^c_B h {H \bar H \over M^2_V}
\left[
1+
\left(H \bar H \over M^2_V \right)+
\left(H \bar H \over M^2_V \right)^2+\ldots
\right]
\left(\theta \over M_V\right)^{p_{AB}} \label{SRNDMRL} \\
(M_{RR})_{AB} &:&
\left\{
F^c_3 F^c_3+
F^c_A F^c_B
\left[
{H \bar H \over M^2_V}+\ldots
\right]
\right\}
{H H \over M^2_V}
\left(\theta \over M_V\right)^{q_{AB}} \label{SRNDMRR}
\end{eqnarray}

The first term in Eq.~\refeqn{SRNDMRL} is renormalizable, thus it implies
third family Yukawa unification at $M_X$. The second term, which we
shall assume to be present for $AB\ne 33$, on the other hand, is a sum
of non-renormalizable operators.
For the sake of simplicity we will consider that the $H\bar H/M_V^2$
part of $\lambda_{AB}$ that lies outside the square brackets in
Eq.~\refeqn{SRNDMRL} has non-trivial gauge contractions with the 
$F_A F^c_B h$ fields next to it, thereby generating the Clebsch
factors in Table \TabClebsch\ (Appendix A). 
On the other hand, the $(H\bar H/M_V^2)^{1,2}$
factors inside the square brackets will form gauge singlet terms that
will be responsible for the appearance of higher $\delta$ powers in 
the entries of $\lambda_{AB}$.
The $M_{RR}$ matrix, as given by Eq.~\refeqn{SRNDMRR}, depends only on 
non-renormalizable operators because gauge invariance demands
that every combination of $F^c F^c$ fields must be paired with at
least a couple of $HH$ fields. However, we will assume that the only
$n=1$ operator in $M_{RR}$ is placed on the 33 entry. 
All other entries of $M_{RR}$ result from $n=2$ operators.
\footnote{We note that these assumptions about the nature of the Majorana
matrix are unique to the LMA MSW solution. 
The SMA MSW and LOW solutions discussed in Appendices C and D are
characterized by a Majorana matrix filled with $n=1$ operators only.}

We can see from Eqs.~\refeqn{SRNDMRL},\refeqn{SRNDMRR} that the structure
of the Yukawa and Majorana matrices can be decomposed into
a ``vertical'' $\delta$-component and a ``horizontal'' 
$\epsilon$-component. Thus we write :
\begin{equation}
(\lambda_f)_{AB} \sim (\lambda^\delta)_{AB} (\lambda^\epsilon)_{AB}
\qquad\qquad
(M_{RR})_{AB} \sim (M_{RR}^\delta)_{AB} (M_{RR}^\epsilon)_{AB}
\label{YukMrrLMA}
\end{equation}
The hierarchies of $\lambda^\epsilon$ and $M_{RR}^\epsilon$ 
are fixed by the choice of the $U(1)_X$ charges.
Using the results of Ref.~\cite{SFKing1}, we can write the most general form
of the unified $(\lambda^\epsilon)_{AB}$ matrix in the 422 model, 
constrained by the absence of anomalies, 
in terms of only four independent parameters 
$\bar X_{F_1}$, $\bar X_{F_2}$, $\bar X_{F^c_1}$ and $\bar X_{F^c_2}$~:
\footnote{In \cite{SFKing1} these charges were called $\alpha$,
$\beta$, $\gamma$, $\delta$.
Roughly, this corresponds to choosing a basis of charges, 
that has $\bar X_{F_3}=\bar X_{F^c_3}=\bar X_{h}=0$.}
\smallskip
\begin{equation}
\lambda^\epsilon =
\left(\matrix{
\epsilon^{|\bar X_{F_1}+\bar X_{F^c_1}|} & 
\epsilon^{|\bar X_{F_1}+\bar X_{F^c_2}|} & 
\epsilon^{|\bar X_{F_1}|} \cr
\epsilon^{|\bar X_{F_2}+\bar X_{F^c_1}|} & 
\epsilon^{|\bar X_{F_2}+\bar X_{F^c_2}|} & 
\epsilon^{|\bar X_{F_2}|} \cr
\epsilon^{|\bar X_{F^c_1}|} & 
\epsilon^{|\bar X_{F^c_2}|} & 
1 \cr}\right)
\label{lambdaepsilonLMA}
\end{equation}
\smallskip
From the equation above it is easy to see that the
values of $\bar X_{F_2}$, $\bar X_{F^c_2}$, $\bar X_{F_1}$ 
and $\bar X_{F^c_1}$
are closely related with the large neutrino $\theta_{23}$ angle,
the second family Yukawa couplings, the $V_{12}$ CKM
angle, and the masses of the lightest fermions respectively.
In the first row of Table ~\TabChargesLMA\ we list our choices for
the $\bar X$'s parameters which we will, from now on, refer to as 
$U(1)_{\bar X}$ charges. In the second row we indicate the values 
of the physical (anomaly free) $U(1)_X=U(1)_{FD}+U(1)_{FI}$ 
charges of the fields of our model.
In the third and forth row we list the values of the family dependent (traceless)
and family independent (unphysical) charges that sum up to give $U(1)_X$.

\vbox{
\begin{center}
\begin{tabular}{lrccccccccc}
\multicolumn{11}{c}{TABLE \TabChargesLMA} \cr
\noalign{\medskip}
\noalign{\hrule height\rulerheight}
\noalign{\smallskip}
\noalign{\hrule height\rulerheight}
\noalign{\medskip}
& &
$X_{F_1}$ & $X_{F_2}$ & $X_{F_3}$ &
$X_{F^c_1}$ & $X_{F^c_2}$ & $X_{F^c_3}$ &
$X_h$ & $X_H$ & $X_{\bar H}$ \\
\noalign{\medskip}
\noalign{\hrule height\rulerheight}
\noalign{\medskip}
$U(1)_{\bar X}$ & : &
$1$ & $\phantom{-}0$ & $\phantom{-}0$ &
$4$ & $2$ & $\phantom{-} 0$ &
$\phantom{-}0$ & $\phantom{-}0$ & $\phantom{-} 0$ \\
\noalign{\smallskip}
$U(1)_X$ & : &
${11 \over 6}$ & $\phantom{-}{5 \over 6}$ & $\phantom{-} {5 \over 6}$ &
${19 \over 6}$ & ${7 \over 6}$ & $-{5 \over 6}$ &
$\phantom{-}0$            & $\phantom{-}{5 \over 6}$ & $-{5 \over 6}$ \\
\noalign{\smallskip}
$U(1)_{FD}$ & : &
${2 \over 3}$ & $-{1 \over 3}$ & $-{1 \over 3}$ &
$2$ & $0$ & $-2$ &
$\phantom{-} {7 \over 3}$ & $\phantom{-}2$ & $-2$ \\
\noalign{\smallskip}
$U(1)_{FI}$ & : &
$ {7 \over 6}$ & $\phantom{-}{7 \over 6}$ & $\phantom{-} {7 \over 6}$ &
$ {7 \over 6}$ & $ {7 \over 6}$ & $\phantom{-}{7 \over 6}$ &
$-{7 \over 3}$ & $-{7 \over 6}$ & $\phantom{-}{7 \over 6}$ \\
\noalign{\medskip}
\noalign{\hrule height\rulerheight}
\noalign{\smallskip}
\noalign{\hrule height\rulerheight}
\end{tabular}
\end{center}

{\narrower\narrower\footnotesize\noindent
{TABLE \TabChargesLMA.} List of $U(1)$ flavour charges that determine
the family structure of the Yukawa and neutrino Majorana matrices 
of the LMA model. The first set, indicated by $U(1)_{\bar X}$, 
refers to the values of the $\bar X$ parameters that determine the
hierarchy of $\lambda^\epsilon$ in Eq.~\refeqn{lambdaepsilonLMA}.
The second set, $U(1)_X = U(1)_{FD}+U(1)_{FI}$, corresponds to the
anomaly free physical flavour charges of the model.
The third set $U(1)_{FD}$ indicates the charges of the family
dependent (traceless) component of $U(1)_X$ and $U(1)_{FI}$ 
refers to the family independent component of $U(1)_X$. 
\par\bigskip}}
\noindent
We note that the $U(1)_{\bar X}$ and $U(1)_X$ charges are
``equivalent'' in the sense that they determine equal family
structures for the Yukawa and neutrino Majorana matrices.~
\footnote{However, only the $U(1)_X$ symmetry is anomaly free.}

The charges in Table~\TabChargesLMA\ fix the $\epsilon$-structure
of $\lambda^\epsilon$ and $M_{RR}^\epsilon$ to be :
\begin{equation}
\lambda^\epsilon\sim\left(\matrix{
\epsilon^5 & \epsilon^3 & \epsilon   \cr
\epsilon^4 & \epsilon^2 & 1 \cr
\epsilon^4 & \epsilon^2 & 1 \cr }\right) \qquad\qquad
M_{RR}^\epsilon\sim\left(\matrix{
\epsilon^8 & \epsilon^6 & \epsilon^4 \cr
\epsilon^6 & \epsilon^4 & \epsilon^2 \cr
\epsilon^4 & \epsilon^2 & 1\cr}\right)
\label{SRNDmRLmRReps}
\end{equation}
Comparing $\lambda^\epsilon$ above with the hierarchy of the Yukawa couplings
listed in Eqs.~\refeqn{SRNDUPdata}-\refeqn{SRNDCKMdata}
we see that, the $U(1)_X$ symmetry (by itself) can not 
explain the pattern of all fermion masses and mixing angles. 
For example, although the symmetry allows a large 23 entry
suitable for generating large 23 mixing from the neutrino Yukawa
matrix, it also allows similarly large 23 entries in the charged
lepton and quark Yukawa matrices which are not welcome.
In order to overcome this we shall assume that although
a renormalizable operator in the 23 position 
is allowed by the $U(1)_X$ symmetry, it is
forbidden by some unspecified string symmetry which however allows
a 23 operator containing one factor of $(H\bar{H})$.
We shall further select a 23 operator which 
will involve a Clebsch factor of zero
for the charged lepton and quark entries, 
with only its neutrino component having a non-zero contribution,
thereby generating a large 23 mixing from the neutrino sector,
with only small 23 mixing in the charged lepton and quark sectors
arising from operators containing higher powers of $(H\bar{H})^n$,
with $n>1$.
The existence of such operators with ``Clebsch zeros'' is clearly
crucial for the success of our approach. 

In general, we shall show that by a suitable choice of non-renormalizable
operators, which determine the $\lambda^\delta$ ``vertical'' structure
of $\lambda_f$, we can obtain a successful description of
all quark and lepton masses and mixing angles.
For example, let us consider $\lambda^\delta$ and $M^\delta_{RR}$
given by the following operator matrices :
\begin{equation}
\lambda^\delta \sim \left(\matrix{
{\cal O}^R+{\cal O}^{\prime\prime V} &
{\cal O}^J+{\cal O}^{\prime Q} &
{\cal O}^g+{\cal O}^{\prime f} \cr
{\cal O}^G+{\cal O}^{\prime\prime K} &
{\cal O}^W+{\cal O}^{\prime H} &
{\cal O}^I+{\cal O}^{\prime W} \cr
{\cal O}^R+{\cal O}^{\prime\prime V} &
{\cal O}^M+{\cal O}^{\prime K} &
1 }\right)
\qquad
M_{RR}^\delta \sim 
\left(\matrix{
{\cal O} & {\cal O} & {\cal O} \cr
{\cal O} & {\cal O} & {\cal O} \cr
{\cal O} & {\cal O} & 1
}\right)
\label{SRNDmRLopMtr}
\end{equation}
where ${\cal O}$, ${\cal O}^{\prime}$ and ${\cal O}^{\prime\prime}$
are $n=1$, $n=2$ 
and (very small) $n=3$ operators respectively~
\footnote{The $n=3$ operators can, to a very good approximation, 
be neglected. Their inclusion here serves only to fill the 11, 21, 31 entries
of the $\lambda_{u,\nu}$ Yukawa matrices, thereby ensuring (for example) 
that the up quark is given a very small mass.}
where $n$ is defined in Eq.~\refeqn{SRNDopRL}
and refers to the powers of $(H\bar{H})^n$.
Using Eqs.~\refeqn{SRNDmRLmRReps},\refeqn{SRNDmRLopMtr} 
into Eq.~\refeqn{YukMrrLMA} gives :
\begin{eqnarray}
\lambda_f(M_X) 
&=& 
\left(\matrix{
\hfil x^R_f a_{11} \delta\epsilon^5 &
\hfil x^J_f a_{12} \delta\epsilon^3 &
\hfil x^g_f a_{13} \delta\epsilon^2 \cr
\hfil x^G_f a_{21} \delta\epsilon^4 &
\hfil x^W_f a_{22} \delta\epsilon^2 &
\hfil x^I_f a_{23} \delta\phantom{\epsilon^2} \cr
\hfil x^R_f a_{31} \delta\epsilon^4 &
\hfil x^M_f a_{32} \delta\epsilon^2 &
\hfil a_{33}\phantom{\epsilon^2} \cr
}\right)+ \nonumber \\
& & \cr
& &
\left(\matrix{
0 &
x^Q_f a^\prime_{12} \delta^2\epsilon^3 &
x^f_f a^\prime_{13} \delta^2\epsilon^2 \cr
0 &
x^H_f a^\prime_{22} \delta^2\epsilon^2 &
\hfil x^W_f a^\prime_{23} \delta^2\phantom{\epsilon^2} \cr
0 &
x^K_f a^\prime_{32} \delta^2\epsilon^2 &
0
}\right)+ \nonumber \\
& & \cr
& &
\left(\matrix{
x^V_f a^{\prime\prime}_{11} \delta^3\epsilon^5 &
0 &
0 \cr
x^K_f a^{\prime\prime}_{21} \delta^3\epsilon^4 &
0 &
0 \cr
x^V_f a^{\prime\prime}_{23} \delta^3\epsilon^4 &
0 &
0 
}\right) \label{lfMXLMAan} \\
& & \cr
{\displaystyle {M_{RR}(M_X)\phantom{_{33}} \over M_{RR}(M_X)_{33}}} 
&=& \left(\matrix{
A_{11} \delta\epsilon^8 & A_{12} \delta\epsilon^6 & A_{13} \delta\epsilon^4 \cr
A_{21} \delta\epsilon^6 & A_{22} \delta\epsilon^4 & A_{23} \delta\epsilon^2 \cr
A_{31} \delta\epsilon^4 & A_{32} \delta\epsilon^2 & A_{33} 
\phantom{\delta\epsilon^2} \cr
}\right) \label{MrrMXLMAan}
\end{eqnarray}
where the subscript $f$ stands for any of the $u,d,e,\nu$ indices,
$x^{\cal O}_f$ is the Clebsch of the ${\cal O}$ operator of the $f$-type
fermion, and the $a$'s ($A$'s) are order-one $f$-independent Yukawa
(Majorana) parameters that parameterise $\lambda_f$ ($M_{RR}$). 
The first matrix on the right-hand side of Eq.\ref{lfMXLMAan}
contains the leading $n=1$ operators giving
contributions of order $\delta$, while the second and third matrices 
contain the $n=2$ and $n=3$ operators 
which give contributions of order $\delta^2$ and $\delta^3$ and provide the
leading contributions in the cases where the $n=1$ operators
involve Clebsch zeros.

The effective matrices resulting from Eqs.~\refeqn{lfMXLMAan},
\refeqn{MrrMXLMAan}
are approximately given in Table~\TabYukMXAprLMA.
This table shows an interesting
structure for the Yukawa matrices. 
We find that 
$\lambda_u\sim\lambda^8$, $\lambda_c\sim\lambda^4$ and
$\lambda_d\sim\lambda_e\sim\lambda^6$,
$\lambda_s\sim\lambda_\mu\sim\lambda^3$.
Furthermore the CKM matrix has
$|V_{12}|\sim\lambda$, $|V_{23}|\sim\lambda^2$,
$|V_{13}|\sim\lambda^3$.
Comparing these approximate results with the data in
Eqs.~\refeqn{SRNDUPdata}-\refeqn{SRNDCKMdata} 
we see that only the $\lambda_d$, $\lambda_\mu$
couplings need substantial (one $\lambda$-power) corrections.
On the other hand, the neutrino sector described by 
$\lambda_\nu$ and $M_{RR}$ in Table~\TabYukMXAprLMA\
is clearly dominated by the right-handed tau neutrino and
predicts $(\lambda_{\nu})_{12}\sim(\lambda_\nu)_{22}$ which according
to Eq.~\refeqn{SRNDt12} successfully generates a large
$\theta_{12}$ solar neutrino angle. 
However, the subdominant perturbation to $m_{LL}$ 
in Eq.~\refeqn{SRNDSeeSaw} resulting from 
$\lambda_\nu$ and $M_{RR}$ in Table~\TabYukMXAprLMA\
is too small to correctly predict the neutrino mass ratio
$m_{\nu_2}/m_{\nu_3}\sim\lambda^{1.5}$ required by Eq.~\refeqn{SRNDneutrinoDATA}.
These approximate predictions can be further improved because
Table~\TabYukMXAprLMA\ does not include the numerical effects of the
operator Clebsches and of the order-one $a$,$A$ factors.

\vbox{
\begin{center}
\begin{tabular}{ccccc}
\multicolumn{5}{c}{TABLE \TabYukMXAprLMA} \cr
\noalign{\medskip}
\noalign{\hrule height\rulerheight}
\noalign{\smallskip}
\noalign{\hrule height\rulerheight}
\noalign{\medskip}
$
\lambda_u(M_X)$ & $\sim$ & $\left(\matrix{
\delta^3\epsilon^5 & \delta^2\epsilon^3 & \delta^2\epsilon\phantom{^3} \cr
\delta^3\epsilon^4 & \delta^2\epsilon^2 & \delta^3 \cr
\delta^3\epsilon^4 & \delta^2\epsilon^2 & 1 \cr} \right)$ & $\sim$ & $
\left(\matrix{
\lambda^8 & \lambda^5 & \lambda^3 \cr
\lambda^7 & \lambda^4 & \lambda^3 \cr
\lambda^7 & \lambda^4 &         1 \cr}\right)
$ \\
\noalign{\smallskip}
$
\lambda_d(M_X)$ & $\sim$ & $\left(\matrix{
\phantom{^3}\delta  \epsilon^5 &
\delta^2\epsilon^3 & 
\delta^2\epsilon\phantom{^3} \cr
\phantom{^3}\delta  \epsilon^4 &
\phantom{^3}\delta  \epsilon^2 & 
\delta^2  \cr
\phantom{^3}\delta  \epsilon^4 & 
\phantom{^3}\delta  \epsilon^2 &
1  \cr} \right)$ & $\sim$ & $
\left(\matrix{
\lambda^6 & \lambda^5 & \lambda^3 \cr
\lambda^5 & \lambda^3 & \lambda^2 \cr
\lambda^5 & \lambda^3 &         1 \cr}\right)
$ \\
\noalign{\smallskip}
$
\lambda_e(M_X)$ & $\sim$ & $\left(\matrix{
\phantom{^3}\delta  \epsilon^5 &
\phantom{^3}\delta  \epsilon^3 & 
\phantom{^3}\delta  \epsilon \phantom{^3} \cr
\phantom{^3}\delta  \epsilon^4 &
\phantom{^3}\delta  \epsilon^2 & 
\delta^2           \cr
\phantom{^3}\delta  \epsilon^4 &
\phantom{^3}\delta  \epsilon^2 &
1                  \cr} \right)$ & $\sim$ & $
\left(\matrix{
\lambda^6 & \lambda^4 & \lambda^2 \cr
\lambda^5 & \lambda^3 & \lambda^2 \cr
\lambda^5 & \lambda^3 &      1 \cr}\right)
$ \\
\noalign{\smallskip}
$
\lambda_\nu(M_X)$ & $\sim$ & $\left(\matrix{
\delta^3\epsilon^5 &
\phantom{^3}\delta\epsilon^3 & 
\phantom{^3}\delta\epsilon \phantom{^3} \cr
\delta^3  \epsilon^4 &
\delta^2  \epsilon^2 & 
\delta           \cr
\delta^3  \epsilon^4 &
\delta^2  \epsilon^2 &
1                  \cr}\right)$ & $\sim$ & $
\left(\matrix{
\lambda^8 & \lambda^4 & \lambda^2 \cr
\lambda^7 & \lambda^4 & \lambda\phantom{1} \cr
\lambda^7 & \lambda^4 & 1 \cr}\right)
$ \\
\noalign{\smallskip}
$
M_{RR}(M_X)$ & $\sim$ & $\left(\matrix{
\delta  \epsilon^8 &
\delta  \epsilon^6 &
\delta  \epsilon^4 \cr
\delta  \epsilon^6 & 
\delta  \epsilon^4 & 
\delta  \epsilon^2 \cr
\delta  \epsilon^4 &
\delta  \epsilon^2 & 1 \cr}\right)$ & $\sim$ & $
\left(\matrix{
\lambda^9 & \lambda^7 & \lambda^5 \cr
\lambda^7 & \lambda^5 & \lambda^3 \cr
\lambda^5 & \lambda^3 &         1 \cr}\right)
$ \\
\noalign{\medskip}
\noalign{\hrule height\rulerheight}
\noalign{\smallskip}
\noalign{\hrule height\rulerheight}
\end{tabular}
\end{center}

{\narrower\narrower\footnotesize\noindent
{TABLE \TabYukMXAprLMA.}
Approximate structure of the Yukawa and neutrino Majorana matrices
resulting from Eqs.~\refeqn{lfMXLMAan},\refeqn{MrrMXLMAan} when the
numerical effect of the Clebsch and of the order-one $a$,$A$ 
parameters is neglected.
\par\bigskip}}

The success of our model (in the SM sector) 
depends on the ability to find suitable
solutions for the $a$'s in 
Eq.~\refeqn{lfMXLMAan} which simultaneously can
account for all the hierarchies in
Eqs.~\refeqn{SRNDUPdata}-\refeqn{SRNDCKMdata}.
Generally we will require that
$0.5 < |a_{AB}|, |a^\prime_{AB}|, |a^{\prime\prime}_{AB}| < 2.0$
for all $A,B=1,2,3$.
At first, it looks that such solution is trivial since
Eq.~\refeqn{lfMXLMAan} depends on 16 parameters,
\footnote{We note that $a_{33}$ is fixed by quadruple Yukawa unification at $M_X$,
{\it i.e.} $\lambda_t=\lambda_b=\lambda_\tau=\lambda_{\nu_\tau}$.}
while Eqs.~\refeqn{SRNDUPdata}-\refeqn{SRNDCKMdata} is a set of 9 constraints
(on the first and second family Yukawa couplings and CKM entries.)
However, we should not forget that
${\cal O} \gg {\cal O}^\prime \gg {\cal O}^{\prime\prime}$ 
and that the CKM matrix constrains only on the 12,13 and 23 entries of $\lambda_f$.
As a consequence, we find that the parameters in 
Eqs.~\refeqn{SRNDUPdata}-\refeqn{SRNDCKMdata} are mainly
sensitive to $a_{22}\leftrightarrow\lambda_{s,\mu}$,
$a_{11}\leftrightarrow\lambda_{d,e}$, an independent combination of 
$(a^{\prime\prime}_{11},a^{\prime\prime}_{21},
  a^{\prime\prime}_{31})\leftrightarrow\lambda_u$,
$a'_{22}\leftrightarrow\lambda_c$ and $a'_{12}\leftrightarrow V_{12}$,
$a'_{23}\leftrightarrow V_{23}$, $a'_{13}\leftrightarrow  V_{13}$, which allows
two predictions to be made --- $\lambda_d$ and $\lambda_s$.
Thus we fitted the $a_{22}$,$a_{11}$,$a^{\prime\prime}_{11}$,
$a'_{22}$,$a'_{12}$,$a'_{23}$,$a'_{13}$
dependence of $\lambda_f$
\footnote{Fixing all other $a$'s to be one.} 
to the $\lambda_\mu$, $\lambda_e$, $\lambda_u$, $\lambda_c$ and
$V_{12}$, $V_{23}$, $V_{13}$ experimental constraints in 
Eqs.~\refeqn{SRNDUPdata}-\refeqn{SRNDCKMdata}. 
The results are shown in Table~\TabaAValuesLMA.
\footnote{The values of the $A$ parameters in Eq.~\refeqn{MrrMXLMAan}
are not constrained by the experimental data, thus we chose them to be
``arbitrary'' numbers in the $0.5 < A_{AB} < 2.0$ range.}

\vbox{
\begin{center}
\begin{tabular}{ccc}
\multicolumn{3}{c}{TABLE \TabaAValuesLMA} \cr
\noalign{\medskip}
\noalign{\hrule height\rulerheight}
\noalign{\smallskip}
\noalign{\hrule height\rulerheight}
\noalign{\medskip}
$
a$ & $=$ & $
\left(\matrix{
-1.285 &
\phantom{-}
 1.000 &
\phantom{-}
 1.000 \cr
\phantom{-}
 1.000 &
-1.420 &
\phantom{-}
 1.000 \cr
-1.000 &
\phantom{-}
 1.000 &
\phantom{-}
 0.677  
}\right)$ \\
\noalign{\smallskip}
$
a^{\prime}$ & $=$ & $
\left(\matrix{
\phantom{-}
 0.000 &
\phantom{-}
 0.700 &
\phantom{-}
 1.015 \cr
\phantom{-}
 0.000 &
\phantom{-}
 0.782 &
\phantom{-}
 0.705 \cr
\phantom{-}
 0.000 &
-1.000 &
\phantom{-}
 0.000
}\right)$ \\
\noalign{\smallskip}
$
a^{\prime\prime}$ & $=$ & $
\left(\matrix{
\phantom{-}
 0.907 &
\phantom{-}
 0.000 &
\phantom{-}
 0.000 \cr
\phantom{-}
 1.000 &
\phantom{-}
 0.000 &
\phantom{-}
 0.000 \cr
\phantom{-}
 1.000 &
\phantom{-}
 0.000 &
\phantom{-}
 0.000
}\right)$ \\
\noalign{\smallskip}
$
A$ & $=$ & $
\left(\matrix{
\phantom{-}
1.071 &
\phantom{-}
0.733 &
\phantom{-}
0.653 \cr
\phantom{-}
0.733 &
\phantom{-}
1.072 &
\phantom{-}
0.567 \cr
\phantom{-}
0.653 &
\phantom{-}
0.576 &
\phantom{-}
1.000 
}\right)$ \\
\noalign{\medskip}
\noalign{\hrule height\rulerheight}
\noalign{\smallskip}
\noalign{\hrule height\rulerheight}
\end{tabular}
\end{center}

{\narrower\narrower\footnotesize\noindent
{TABLE \TabaAValuesLMA.} 
Numerical values of the order-one $a$,$A$ parameters 
that parameterise the Yukawa and neutrino Majorana matrices in
Eqs.~\refeqn{lfMXLMAan},\refeqn{MrrMXLMAan}.
\par\bigskip}}

Thus, using Eq.~\refeqn{lfMXLMAan} with the $a$'s of
Table~\TabaAValuesLMA\ and the Clebsch factors of Table~\TabClebsch\ (Appendix A) 
we get the numerical results for $\lambda_f(M_X)$ shown in Table~\TabYukMXnumLMA.
In Table~\TabYukMXepsLMA\ we present the results of
Table~\TabYukMXnumLMA\ expanded in powers of $\delta=\epsilon=\lambda=0.22$.

\vbox{
\vbox{
\begin{center}
\begin{tabular}{ccc}
\multicolumn{3}{c}{TABLE \TabYukMXnumLMA} \cr
\noalign{\medskip}
\noalign{\hrule height\rulerheight}
\noalign{\smallskip}
\noalign{\hrule height\rulerheight}
\noalign{\medskip}
$
\lambda_u(M_X) $ & $=$ & $\left(\matrix{
\phantom{-}
7.034\times 10^{-6} &
\phantom{-}
4.079\times 10^{-4} &
\phantom{-}
4.324\times 10^{-3} \cr
\phantom{-}
3.991\times 10^{-5} &
\phantom{-}
1.466\times 10^{-3} &
\phantom{-}
0.000 \cr
\phantom{-}
3.528\times 10^{-5} &
-
3.748\times 10^{-3} &
\phantom{-}
0.677 }\right) 
$ \\
\noalign{\smallskip}
$
\lambda_d(M_X)$ & $=$ & $\left(\matrix{
-
2.331\times 10^{-4} &
-
4.079\times 10^{-4} &
\phantom{-}
8.648\times 10^{-3} \cr
\phantom{-}
4.609\times 10^{-4} &
-
8.827\times 10^{-3} &
\phantom{-}
2.157\times 10^{-2} \cr
-
8.246\times 10^{-4} &
\phantom{-}
1.506\times 10^{-2} &
\phantom{-}
0.677 }\right)
$ \\
\noalign{\smallskip}
$
\lambda_e(M_X)$ & $=$ & $\left(\matrix{
-
1.748\times 10^{-4} &
\phantom{-}
3.884\times 10^{-3} &
\phantom{-}
 8.574\times 10^{-2} \cr
\phantom{-}
9.219\times 10^{-4} &
\phantom{-}
 3.015\times 10^{-2} &
-
 6.472\times 10^{-2} \cr
-
6.184\times 10^{-4} &
\phantom{-}
 1.501\times 10^{-2} &
\phantom{-}
0.677}\right)
$ \\
\noalign{\smallskip}
$
\lambda_\nu(M_X)$ & $=$ & $\left(\matrix{
\phantom{-}
 7.034\times 10^{-6} &
\phantom{-}
 2.401\times 10^{-3} &
\phantom{-}
 7.710\times 10^{-2} \cr
\phantom{-}
 2.993\times 10^{-5} &
\phantom{-}
 2.932\times 10^{-3} &
\phantom{-}
 0.440 \cr
\phantom{-}
 3.528\times 10^{-5} &
-
 2.811\times 10^{-3} &
\phantom{-}
 0.677}\right)
$ \\
\noalign{\smallskip}
$
M_{RR}(M_X)$ & $=$ & $\left(\matrix{
\phantom{-}
 3.991 \times 10^{ 8}\phantom{^1} &
\phantom{-}
 5.652 \times 10^{ 9}\phantom{^1} &
\phantom{-}
 1.040 \times 10^{11} \cr
\phantom{-}
 5.652 \times 10^{ 9}\phantom{^1} &
\phantom{-}
 1.706 \times 10^{11} &
\phantom{-}
 1.866 \times 10^{12} \cr
\phantom{-}
 1.040 \times 10^{11} &
\phantom{-}
 1.866 \times 10^{12} &
\phantom{-}
 3.090 \times 10^{14} \cr
}\right)
$ \\
\noalign{\medskip}
\noalign{\hrule height\rulerheight}
\noalign{\smallskip}
\noalign{\hrule height\rulerheight}
\end{tabular}
\end{center}
\vskip -10pt
{\narrower\narrower\footnotesize\noindent
{TABLE \TabYukMXnumLMA.} 
Numerical values for the entries of the Yukawa 
($\lambda_u$,$\lambda_d$,$\lambda_e$,$\lambda_\nu$)
and neutrino Majorana ($M_{RR}$) matrices at the unification scale $M_X$ 
($M_{RR}$ is given in GeV mass units.)
\par}}
\vskip -15pt
\vbox{
\begin{center}
\begin{tabular}{ccc}
\multicolumn{3}{c}{TABLE \TabYukMXepsLMA} \cr
\noalign{\medskip}
\noalign{\hrule height\rulerheight}
\noalign{\smallskip}
\noalign{\hrule height\rulerheight}
\noalign{\medskip}
$
\lambda_u(M_X)$ & $\sim$ & $\left(\matrix{
\phantom{-}
\lambda^{7.836} &
\phantom{-}
\lambda^{5.154} &
\phantom{-}
\lambda^{3.595} \cr
\phantom{-}
\lambda^{6.690} &
\phantom{-}
\lambda^{4.309} &
0               \cr
\phantom{-}
\lambda^{6.771} &
-
\lambda^{3.690} &
\phantom{-}
\lambda^{0.257} \cr} \right)
$ \\
\noalign{\smallskip}
$
\lambda_d(M_X)$ & $\sim$ & $\left(\matrix{
-
\lambda^{5.524} &
-
\lambda^{5.154} &
\phantom{-}
\lambda^{3.137} \cr
\phantom{-}
\lambda^{5.074} &
-
\lambda^{3.124} &
\phantom{-}
\lambda^{2.534} \cr
-
\lambda^{4.690} &
\phantom{-}
\lambda^{2.771} &
\phantom{-}
\lambda^{0.257} \cr} \right)
$ \\
\noalign{\smallskip}
$
\lambda_e(M_X)$ & $\sim$ & $\left(\matrix{
-
\lambda^{5.714} &
\phantom{-}
\lambda^{3.666} &
\phantom{-}
\lambda^{1.622} \cr
\phantom{-}
\lambda^{4.616} &
\phantom{-}
\lambda^{2.313} &
-
\lambda^{1.808} \cr
-
\lambda^{4.880} &
\phantom{-}
\lambda^{2.771} &
\phantom{-}
\lambda^{0.257} \cr} \right)
$ \\
\noalign{\smallskip}
$
\lambda_\nu(M_X)$ & $\sim$ & $\left(\matrix{
\phantom{-}
\lambda^{7.836} &
\phantom{-}
\lambda^{3.984} &
\phantom{-}
\lambda^{1.693} \cr
\phantom{-}
\lambda^{6.880} &
\phantom{-}
\lambda^{3.852} &
\phantom{-}
\lambda^{0.542} \cr
\phantom{-}
\lambda^{6.771} &
-
\lambda^{3.880} &
\phantom{-}
\lambda^{0.257} \cr} \right)
$ \\
\noalign{\smallskip}
$
{\displaystyle{M_{RR}(M_X)\phantom{_{33}} \over 
               M_{RR}(M_X)_{33}}}$ & $\sim$ & $\left(\matrix{
\phantom{-}
\lambda^{8.955} &
\phantom{-}
\lambda^{7.205} &
\phantom{-}
\lambda^{5.281} \cr
\phantom{-}
\lambda^{7.205} &
\phantom{-}
\lambda^{4.954} &
\phantom{-}
\lambda^{3.375} \cr
\phantom{-}
\lambda^{5.281} &
\phantom{-}
\lambda^{3.375} &
1 \cr} \right)
$ \\
\noalign{\medskip}
\noalign{\hrule height\rulerheight}
\noalign{\smallskip}
\noalign{\hrule height\rulerheight}
\end{tabular}
\end{center}
\vskip -10pt
{\narrower\narrower\footnotesize\noindent
{TABLE \TabYukMXepsLMA.} 
Values for the entries of the Yukawa 
($\lambda_u$,$\lambda_d$,$\lambda_e$,$\lambda_\nu$)
and neutrino Majorana ($M_{RR}$) matrices at the unification scale $M_X$ 
as given in Table~\TabYukMXnumLMA\ but expanded in powers of 
$\delta=\epsilon=\lambda=0.22$
(Note $M_{RR}(M_X)_{33}=3.090\times 10^{14} {\rm\ GeV}$.)
\par\bigskip}}}

We can analyse the effect of the operator Clebsches by comparing 
Table~\TabYukMXAprLMA\ against Table~\TabYukMXepsLMA.
We see that the ${\cal O}^W$ operator
\footnote{That has Clebsches $x^W_e = -3 x^W_d$}
in the 22 entry of $\lambda_f$ 
split $(\lambda_d)_{22}=(\lambda_e)_{22}\sim\lambda^3$ in 
Table~\TabYukMXAprLMA\ into
$|(\lambda_d)_{22}|=\lambda^{3.124}$ and $(\lambda_e)_{22}=\lambda^{2.313}$
in Table~\TabYukMXepsLMA, thus allowing for a
proper GJ ratio $\lambda_\mu / \lambda_s\sim 3$ at $M_X$.
Similarly, the operators in the 12 block have allowed for a more
appropriate $\lambda_d / \lambda_e$ Yukawa ratio. Numerically we have
the following predictions for the lightest eigenvalues of the
down-Yukawa matrix at $M_X$ : $\lambda_d=\lambda^{5.469}$ and
$\lambda_s=\lambda^{3.087}$.

The effect of the Clebsch factors also modified the neutrino Yukawa
matrix $\lambda_\nu(M_X)$ in Table~\TabYukMXAprLMA. 
Due to the ${\cal O}^I$ operator in the 23
position of $\lambda_f$, which has a large $x^I_\nu=2$ Clebsch, we are
now able to predict a large $\theta_{23}$ atmospheric neutrino
mixing angle. Indeed using Eq.~\refeqn{SRNDRotAngles} 
we can roughly estimate that $\tan\theta_{23}\sim 0.44/0.68=0.65$ implying 
$\sin^2(2\theta_{23}) \sim 0.83$.

It is interesting to check that $\lambda_\nu$ and $M_{RR}$ 
in Table~\TabYukMXepsLMA\ do lead to a $m_{LL}$ matrix dominated by
the right-handed tau neutrino. 
As a result of the small mixing angles of $M_{RR}$
it is convenient to work in a basis where $M_{RR}$ is
diagonal. Furthermore it is practical to scale $\lambda_\nu$ 
\footnote{
$(\lambda_\nu)_{AB} \to (\lambda_\nu)_{AB} / k$
with $k = {1\over 2} [ (\lambda_\nu)_{23}+(\lambda_\nu)_{33}] = \lambda^{0.384}$.} 
such that the 23 and 33 entries are 
$(\lambda_\nu)_{23}\sim(\lambda_\nu)_{33}\sim 1$,
and approximate the normalized entries to 
(semi-)integer $\lambda$-powers. Thus using :
\begin{equation}
M_{RR}\sim\left(\matrix{
\lambda^{9.0} & 0 & 0   \cr
0 & \lambda^{5.0} & 0 \cr
0 & 0 & 1 }\right)
\qquad
m_{RL}\sim\left(\matrix{
\lambda^{7.5} & \lambda^{3.5} & \lambda^{1.5} \cr
\lambda^{6.5} & \lambda^{3.5} & 1 \cr
\lambda^{6.5} & \lambda^{3.5} & 1 }\right) 
\end{equation}
into Eq.~\refeqn{SRNDSeeSaw} gives :
\begin{equation}
m_{LL} \sim \left(\matrix{
\lambda^{3.0}+\lambda^{2.0}+\lambda^{6.0} &
\lambda^{1.5}+\lambda^{2.0}+\lambda^{5.0} &
\lambda^{1.5}+\lambda^{2.0}+\lambda^{5.0} \cr
\lambda^{1.5}+\lambda^{2.0}+\lambda^{5.0} &
\hfill 1+\lambda^{2.0}+\lambda^{4.0} &
\hfill 1+\lambda^{2.0}+\lambda^{4.0} \cr
\lambda^{1.5}+\lambda^{2.0}+\lambda^{5.0} &
\hfill 1+\lambda^{2.0}+\lambda^{4.0} &
\hfill 1+\lambda^{2.0}+\lambda^{4.0} }\right)
\label{mLLGUTapr}
\end{equation}
where the first (second and third) term in each entry corresponds to
the third (second and first) family neutrino contribution 
$\nu^c_\tau$ ($\nu^c_\mu$ and $\nu^c_e$) coming from $M_{RR}$.
Clearly Eq.~\refeqn{mLLGUTapr} shows that 
even though, in this case, $\nu^c_\tau$ is the heaviest right-handed
neutrino, it nevertheless dominates the 23 block,
and that the sub-dominant contribution from $\nu^c_\mu$ induces
$\lambda^2$ perturbations in $m_{LL}$ that are 
compatible with the $m_{\nu_2} / m_{\nu_3}$ mass ratio.

Using the MSSM RGEs adapted and properly extended to take into
account the presence and successive decoupling of the right-handed
neutrinos between the $Q=M_X$ and $Q=M_Z$ scales (see Appendix B)
we find that the Yukawa and neutrino Majorana matrices
at low energy are given by Table~\TabYukMznumLMA.

\vbox{
\begin{center}
\begin{tabular}{ccc}
\multicolumn{3}{c}{TABLE \TabYukMznumLMA} \cr
\noalign{\medskip}
\noalign{\hrule height\rulerheight}
\noalign{\smallskip}
\noalign{\hrule height\rulerheight}
\noalign{\medskip}
$
\lambda_u(M_Z) $ & $=$ & $\left(\matrix{
\phantom{-}
1.478\times 10^{-5} &
\phantom{-}
8.920\times 10^{-4} &
\phantom{-}
5.058\times 10^{-3} \cr
\phantom{-}
8.484\times 10^{-5} &
\phantom{-}
3.143\times 10^{-3} &
-
3.553\times 10^{-3} \cr
\phantom{-}
4.687\times 10^{-5} &
-
5.015\times 10^{-3} &
0.905 }\right) 
$ \\
\noalign{\smallskip}
$
\lambda_d(M_Z)$ & $=$ & $\left(\matrix{
-
4.605\times 10^{-4} &
-
9.389\times 10^{-4} &
\phantom{-}
1.175\times 10^{-2} \cr
\phantom{-}
9.379\times 10^{-4} &
-
1.795\times 10^{-2} &
\phantom{-}
3.124\times 10^{-2} \cr
-
1.057\times 10^{-3} &
\phantom{-}
1.934\times 10^{-2} &
\phantom{-}
0.862 }\right)
$ \\
\noalign{\smallskip}
$
\lambda_e(M_Z)$ & $=$ & $\left(\matrix{
-
1.526\times 10^{-4} &
\phantom{-}
3.445\times 10^{-3} &
\phantom{-}
 6.611\times 10^{-2} \cr
\phantom{-}
8.832\times 10^{-4} &
\phantom{-}
 2.931\times 10^{-2} &
-
 5.446\times 10^{-2} \cr
-
4.626\times 10^{-4} &
\phantom{-}
 1.196\times 10^{-2} &
\phantom{-}
0.522}\right)
$ \\
\noalign{\smallskip}
$
\lambda_\nu(M_Z)$ & $=$ & $\left(\matrix{
\phantom{-}
 6.110\times 10^{-6} &
\phantom{-}
 2.271\times 10^{-3} &
\phantom{-}
 6.339\times 10^{-2} \cr
\phantom{-}
 2.699\times 10^{-5} &
\phantom{-}
 2.737\times 10^{-3} &
\phantom{-}
 0.394 \cr
\phantom{-}
 2.911\times 10^{-5} &
-
 2.428\times 10^{-3} &
\phantom{-}
 0.562}\right)
$ \\
\noalign{\smallskip}
$
M_{RR}(M_Z)$ & $=$ & $\left(\matrix{
\phantom{-}
 3.987 \times 10^{ 8}\phantom{^1} &
\phantom{-}
 5.651 \times 10^{ 9}\phantom{^1} &
\phantom{-}
 9.981 \times 10^{10} \cr
\phantom{-}
 5.651 \times 10^{ 9}\phantom{^1} &
\phantom{-}
 1.707 \times 10^{11} &
\phantom{-}
 1.807 \times 10^{12} \cr
\phantom{-}
 9.981 \times 10^{10} &
\phantom{-}
 1.807 \times 10^{12} &
\phantom{-}
 2.879 \times 10^{14} \cr
}\right)
$ \\
\noalign{\medskip}
\noalign{\hrule height\rulerheight}
\noalign{\smallskip}
\noalign{\hrule height\rulerheight}
\end{tabular}
\end{center}

{\narrower\narrower\footnotesize\noindent
{TABLE \TabYukMznumLMA.} 
Numerical values for the entries of the Yukawa 
($\lambda_u$,$\lambda_d$,$\lambda_e$,$\lambda_\nu$)
and neutrino Majorana ($M_{RR}$) matrices at the $Q=M_Z$ scale 
($M_{RR}$ is given in GeV mass units.)
\par\bigskip}}
\noindent
Thus, inserting the results of Table~\TabYukMznumLMA\ into
Eq.~\refeqn{SRNDSeeSaw} and Eq.~\refeqn{SRNDCKMMNS}
we get the mass matrix for the left-handed neutrinos $m_{LL}$ 
and the $V^{MNS}$ mixing matrix shown in Table~\TabMllVmnsLMA.
The predictions for the neutrino masses and squared mass splittings
are shown in Table~\TabNeutMassLMA.
In Table~\TabAnglesLMA\ we examine how the neutrino mixing angles
evolve between the unification
$Q=M_X\sim 3\times 10^{16} {\rm\ GeV}$,
the right-handed tau neutrino mass $Q=M_{\nu_3}\sim 3\times 10^{14} {\rm\ GeV}$
and the $M_Z$ scale. 
We see that the effect of the radiative corrections has increased the
magnitude of $\sin\theta_{12}$, $\sin\theta_{23}$ and
$\sin\theta_{13}$ by 2.5\%, 6.4\% and 2.4\% respectively.
These corrections agree with the results found in Ref.~\cite{KiSi}.
Finally, we present in Table~\TabmdmsLMA\ the predictions
for the down and strange quark masses.

We would like to conclude this section by noting that the predictions
for the neutrino parameters, in particular for the neutrino 
$\Delta m_{12}^2$ squared mass splitting,
should be taken carefully.
Generally we expect at least 20\% (theoretical) errors in the quoted
values which, for example, arise from our inability
to fix order-one factors in the entries of $M_{RR}(M_X)$.

\vbox{
\begin{center}
\begin{tabular}{ccl}
\multicolumn{3}{c}{TABLE \TabMllVmnsLMA} \cr
\noalign{\medskip}
\noalign{\hrule height\rulerheight}
\noalign{\smallskip}
\noalign{\hrule height\rulerheight}
\noalign{\medskip}
$
{\displaystyle {m_{LL}(M_Z)\phantom{_{33}} \over m_{LL}(M_Z)_{33}}}$ & $=$ & $ 
\left(\matrix{
4.792\times 10^{-2} &
1.007\times 10^{-1} &
3.458\times 10^{-2} \cr
1.007\times 10^{-1} &
4.610\times 10^{-1} &
5.659\times 10^{-1} \cr
3.458\times 10^{-2} &
5.659\times 10^{-1} &
1 }\right) $ \\
\noalign{\smallskip}
$
V^{MNS}(M_Z)$ & $=$ & $\left(\matrix{
\phantom{-} 0.8290  & 
\phantom{-} 0.5532  &
           -0.0819 \cr
           -0.3948  &
\phantom{-} 0.6827  &
\phantom{-} 0.6149 \cr
\phantom{-} 0.3961 &
           -0.4774 &
\phantom{-} 0.7843 }\right)$ \\

\noalign{\medskip}
\noalign{\hrule height\rulerheight}
\noalign{\smallskip}
\noalign{\hrule height\rulerheight}
\end{tabular}
\end{center}

{\narrower\narrower\footnotesize\noindent
{TABLE \TabMllVmnsLMA.}
Predicted values for the left-handed neutrino mass matrix $m_{LL}(M_Z)$
in units of $m_{LL}(M_Z)_{33} = 3.954\times 10^{-3} {\rm\ eV}$ 
and for the MNS neutrino mixing matrix $V^{MNS}(M_Z)$.
\par\bigskip}}

\vbox{
\begin{center}
\begin{tabular}{c}
\multicolumn{1}{c}{TABLE \TabNeutMassLMA} \cr
\noalign{\medskip}
\noalign{\hrule height\rulerheight}
\noalign{\smallskip}
\noalign{\hrule height\rulerheight}
\noalign{\medskip}
$
m_{\nu_1} = 4.84\times 10^{-8} {\rm\ eV}         \quad
m_{\nu_2} = 5.79\times 10^{-3} {\rm\ eV}         \quad
m_{\nu_3} = 5.39\times 10^{-2} {\rm\ eV}               
$ \\
\noalign{\medskip}
$
\Delta m_{12}^2 = 3.35\times 10^{-5} {\rm\ eV}^2 \qquad
\Delta m_{23}^2 = 2.87\times 10^{-3} {\rm\ eV}^2 
$ \\
\noalign{\medskip}
\noalign{\hrule height\rulerheight}
\noalign{\smallskip}
\noalign{\hrule height\rulerheight}
\end{tabular}
\end{center}

{\narrower\narrower\footnotesize\noindent
{TABLE \TabNeutMassLMA.}
Predicted values for the left-handed neutrino masses $m_{\nu_{1,2,3}}$
and squared mass splittings 
$\Delta m^2_{12} = |m_{\nu_2}^2-m_{\nu_1}^2|$,
$\Delta m^2_{23} = |m_{\nu_3}^2-m_{\nu_2}^2|$.
\par\bigskip}}

\vbox{
\begin{center}
\begin{tabular}{rrr}
\multicolumn{3}{c}{TABLE \TabAnglesLMA} \cr
\noalign{\medskip}
\noalign{\hrule height\rulerheight}
\noalign{\smallskip}
\noalign{\hrule height\rulerheight}
\noalign{\medskip}
\multicolumn{3}{c}{$Q = M_X\sim 3\times 10^{16} {\rm\ GeV}$} \cr
$\sin\theta_{12} = 0.541$ &
$\sin\theta_{23} = 0.578$ &
$\sin\theta_{13} =-0.080$ \\
$\sin^2(2\theta_{12}) = 0.828$ &
$\sin^2(2\theta_{23}) = 0.890$ &
$\sin^2(2\theta_{13}) = \phantom{-}0.025$ \\
\noalign{\medskip}
\noalign{\hrule height\rulerheight}
\noalign{\medskip}
\multicolumn{3}{c}{$Q = M_{\nu_3}\sim 3\times 10^{14} {\rm\ GeV}$} \cr
$\sin\theta_{12} = 0.543$ &
$\sin\theta_{23} = 0.590$ &
$\sin\theta_{13} =-0.082$ \\
$\sin^2(2\theta_{12}) = 0.832$ &
$\sin^2(2\theta_{23}) = 0.908$ &
$\sin^2(2\theta_{13}) = \phantom{-}0.027$ \\
\noalign{\medskip}
\noalign{\hrule height\rulerheight}
\noalign{\medskip}
\multicolumn{3}{c}{$Q = M_Z$} \cr
$\sin\theta_{12} = 0.555$ &
$\sin\theta_{23} = 0.617$ &
$\sin\theta_{13} =-0.082$ \\
$\sin^2(2\theta_{12}) = 0.853$ &
$\sin^2(2\theta_{23}) = 0.943$ &
$\sin^2(2\theta_{13}) = \phantom{-}0.027 $ \\
\noalign{\medskip}
\noalign{\hrule height\rulerheight}
\noalign{\smallskip}
\noalign{\hrule height\rulerheight}
\end{tabular}
\end{center}

{\narrower\narrower\footnotesize\noindent
{TABLE \TabAnglesLMA.} 
Running of the neutrino mixing angles at the unification $Q=M_X$, 
the right-handed tau neutrino mass $Q=M_{\nu_3}$ and $Z$ boson mass $Q=M_Z$
energy scales.
\par\bigskip}}

\vbox{
\begin{center}
\begin{tabular}{ll}
\multicolumn{2}{c}{TABLE \TabmdmsLMA} \cr
\noalign{\medskip}
\noalign{\hrule height\rulerheight}
\noalign{\smallskip}
\noalign{\hrule height\rulerheight}
\noalign{\medskip}
$m_d(1 {\rm\ GeV}) = 4.9 {\rm\ MeV}          \quad $ &
$m_s(M_s) = 156 {\rm\ MeV}                   $ \\
\noalign{\medskip}
\noalign{\hrule height\rulerheight}
\noalign{\smallskip}
\noalign{\hrule height\rulerheight}
\end{tabular}
\end{center}

{\narrower\narrower\footnotesize\noindent
{TABLE \TabmdmsLMA.} 
Predictions for the running $\overline{\rm MS}$ masses of the down ($m_d$)
and strange ($m_s$) quarks  at $Q=1 {\rm\ GeV}$ and $Q=M_s$ respectively
($M_s$ indicates the strange pole quark mass.)
\par\bigskip}}

\newpage

\SECTION{IX. CONCLUSION}

We have discussed a theory of all fermion masses and
mixing angles based on a particular string-inspired
{\em minimal} model based on the Pati-Salam group 
$SU(4)\times SU(2)_L \times SU(2)_R$ \cite{PaSa}
supplemented by a gauged $U(1)_X$ family symmetry.
We argued that this gauge group preserves the attractive features
of $SO(10)$ such as predicting three right-handed neutrinos,
and Yukawa unification, while avoiding the doublet-triplet splitting
problem. Although it is not a unified gauge group at the field
theory level, it naturally arises from string constructions and
so in principle may be fully unified with gravity.

Earlier work in collaboration with one of us \cite{AlKiLeLo1}
had already shown that
the model can provide a successful description of the
charged fermion masses and the CKM matrix. The use of the
$U(1)_X$ family symmetry to provide the horizontal mass
splittings combined with the Clebsch factors arising from the $(H\bar{H})^n$
insertion in the operators has already been shown to provide a
powerful approach to the fermion mass spectrum in this model \cite{AlKiLeLo1}.
The present analysis differs from that presented previously 
partly due to the recent refinements in third family Yukawa unification 
\cite{KiOl2}, but mainly due to the recent data from Super-Kamiokande 
which implies that the 23 operator should be allowed by
the $U(1)_X$ family symmetry.
We have therefore extended our previous analysis to 
the atmospheric and solar neutrino masses
and mixing angles, and showed that all three MSW 
solutions to the solar neutrino data may be
accommodated, namely the LMA MSW region discussed
in the main text as well as the LOW MSW and the SMA MSW regions
discussed in the Appendices.

The approach to neutrino masses and mixing angles followed here
makes use of the SRHND mechanism \cite{KingSRND,SFKing1,SFKing2}
in which one of the right-handed neutrinos (the $\nu_{\tau}^c$) gives the
dominant contribution to the 23 block of the light effective
Majorana matrix. This mechanism avoids reliance on accidental
cancellations, and does not rely on excessive magnification
of mixing angles, although a mild enhancement was observed in the
numerical results in agreement with that observed in \cite{KiSi}.
Crucial to the implementation of SRHND in this model is the assumption
that the renormalizable 23 operator is
forbidden by unspecified string selection rules,
and the leading 23 operator contains $(H\bar{H})$ and involves
``Clebsch zeros'', which give a zero contribution
to the charged lepton and quark Yukawa matrices, but a non-zero
contribution to the neutrino Yukawa matrix, thereby allowing
small $V_{cb}$ but large 23 mixing in the lepton sector.

The analysis in this paper is essentially ``bottom-up''.
A particular choice of $U(1)_X$ family symmetry charges 
was used to give the horizontal mass splittings, 
and the vertical mass splittings
were achieved by particular choices of operators corresponding
to different Clebsch factors in the leading contributions to
each entry of the Yukawa matrix.
It would be very nice to understand these choices from the
point of view of a ``top-down'' string construction, such as
the Type I string construction which has recently led to the Pati-Salam
gauge group with three chiral families \cite{ShTy}.
We believe that only by a combination of top-down
and bottom-up approaches (such as that presented here)
will a completely successful string theory of fermion
masses and mixing angles emerge.
We have shown that the recent discovery of neutrino mass by Super-Kamiokande
provides precious information about the flavour structure of
such a future string theory.

\SECTION{ACKNOWLEDGMENTS}
The work of M.O. was supported by JNICT under contract grant :
PRAXIS XXI/BD/ 5536/95.
 
\newpage

\SECTION{APPENDIX A}
\vskip -20pt
\vbox{
\begin{center}
\begin{tabular}{clrrrr}
\multicolumn{6}{c}{TABLE \TabClebsch} \cr
\noalign{\medskip}
\noalign{\hrule height\rulerheight}
\noalign{\smallskip}
\noalign{\hrule height\rulerheight}
\noalign{\medskip}
CLASS & 
${\cal O}$ & 
$x_u$ & 
$x_d$ & 
$x_e$ &
$x_\nu$ \\
\noalign{\medskip}
\noalign{\hrule height\rulerheight}
\noalign{\medskip}
I     & ${ \cal O}^N $& 2.0000 & 0.0000 & 0.0000 & 0.0000  \\
I     & ${ \cal O}^E $& 0.0000 & 2.0000 & 0.0000 & 0.0000  \\
I     & ${ \cal O}^i $& 0.0000 & 0.0000 & 2.0000 & 0.0000  \\
I     & ${ \cal O}^I $& 0.0000 & 0.0000 & 0.0000 & 2.0000  \\
II    & ${ \cal O}^M $& 0.0000 & 1.4142 & 1.4142 & 0.0000  \\
II    & ${ \cal O}^G $& 0.0000 & 0.8944 & 1.7889 & 0.0000  \\
II    & ${ \cal O}^R $& 0.0000 & 1.6000 & 1.2000 & 0.0000  \\
II    & ${ \cal O}^W $& 0.0000 & 0.6325 &-1.8974 & 0.0000  \\
III   & ${ \cal O}^V $& 1.4142 & 0.0000 & 0.0000 & 1.4142  \\
III   & ${ \cal O}^O $& 0.8944 & 0.0000 & 0.0000 & 1.7889  \\
III   & ${ \cal O}^K $& 1.6000 & 0.0000 & 0.0000 & 1.2000  \\
III   & ${ \cal O}^Z $& 0.6325 & 0.0000 & 0.0000 &-1.8974  \\
IV    & ${ \cal O}^J $& 0.0000 & 0.0000 & 1.7889 & 0.8944  \\
IV    & ${ \cal O}^g $& 0.0000 & 0.0000 & 1.4142 & 1.4142  \\
IV    & ${ \cal O}^h $& 0.0000 & 0.0000 &-1.4142 & 1.4142  \\
IV    & ${ \cal O}^j $& 0.0000 & 0.0000 & 0.8944 & 1.7889  \\
V     & ${\cal O}^F $& 1.4142 &-1.4142 & 0.0000 & 0.0000  \\
V     & ${\cal O}^a $& 1.4142 & 1.4142 & 0.0000 & 0.0000  \\
V     & ${\cal O}^b $& 1.7889 & 0.8944 & 0.0000 & 0.0000  \\
V     & ${\cal O}^c $& 0.8944 & 1.7889 & 0.0000 & 0.0000  \\
VI    & ${\cal O}^H $& 0.8000 & 0.4000 & 0.8000 & 1.6000  \\
VI    & ${\cal O}^f $& 0.4000 & 0.8000 & 1.6000 & 0.8000  \\
VI    & ${\cal O}^S $& 0.7155 & 1.4311 & 1.0733 & 0.5367  \\
VI    & ${\cal O}^L $& 1.4311 & 0.7155 & 0.5367 & 1.0733  \\
VI    & ${\cal O}^T $& 0.5657 & 0.2828 & 0.2828 & 0.5657  \\
VI    & ${\cal O}^U $& 0.2828 & 0.5657 & 0.5657 & 0.2828  \\
VI    & ${\cal O}^X $& 0.5657 & 0.2828 &-0.8485 &-1.6971  \\
VI    & ${\cal O}^Y $& 0.2828 & 0.5657 &-1.6971 &-0.8485  \\
VI    & ${\cal O}^D $& 0.4472 &-0.4472 & 1.3416 &-1.3416  \\
VI    & ${\cal O}^e $& 0.6325 &-0.6325 &-1.2649 & 1.2649  \\
VI    & ${\cal O}^B $& 1.0000 &-1.0000 &-1.0000 & 1.0000  \\
VI    & ${\cal O}^Q $& 1.1314 &-1.1314 &-0.8485 & 0.8485  \\
VI    & ${\cal O}^P $& 1.1314 & 1.1314 & 0.8485 & 0.8485  \\
VI    & ${\cal O}^A $& 1.0000 & 1.0000 & 1.0000 & 1.0000  \\
VI    & ${\cal O}^d $& 0.6325 & 0.6325 & 1.2649 & 1.2649  \\
VI    & ${\cal O}^C $& 0.4472 & 0.4472 &-1.3416 &-1.3416  \\
\noalign{\medskip}
\noalign{\hrule height\rulerheight}
\noalign{\smallskip}
\noalign{\hrule height\rulerheight}
\end{tabular}
\end{center}
\vskip -10pt
{\narrower\narrower\footnotesize\noindent
{TABLE \TabClebsch.}
List of Clebsch factors resulting from all possible $n=1$ operators,
as given by Eq.~\refeqn{SRNDopRL}, in the Pati-Salam model.
\par\bigskip}}

\newpage

\SECTION{APPENDIX B}

In this appendix we briefly review some technical issues related to
the presence of the right-handed neutrinos. Firstly we show how the
decoupling of the neutrinos affects the one-loop RGEs for
the Yukawa couplings in the MSSM+$\nu^c$ model :
\begin{eqnarray}
{1 \over 16\pi^2}
{d\lambda_u\over dt} &=&
[ 3{\rm Tr}U+{\rm Tr}N+3U+D-G^u] \lambda_u \label{SRNDrgeU} \\
{1 \over 16\pi^2}
{d\lambda_d\over dt} &=&
[ 3{\rm Tr}D+{\rm Tr}E+3D+U-G^d] \lambda_d \\
{1 \over 16\pi^2}
{d\lambda_e\over dt} &=&
[ 3{\rm Tr}D+{\rm Tr}E+3E+N-G^e] \lambda_e \\
{1 \over 16\pi^2}
{d\lambda_\nu\over dt} &=&
[ 3{\rm Tr}U+{\rm Tr}N+3N+E-G^\nu] \lambda_\nu \label{SRNDrgeN}
\end{eqnarray}
where $t=\ln(Q)$,
\begin{equation}
\matrix{ 
U   = \lambda_u\lambda_u^\dagger \hfill & \phantom{space}
G^u = {26 \over 30} g^2_1+3g^2_2+{16\over 3}g^2_3 \hfill \cr
D   = \lambda_d\lambda_d^\dagger \hfill & \phantom{space}
G^d = {14 \over 30} g^2_1+3g^2_2+{16 \over 3}g^2_3 \hfill \cr
E   = \lambda_e\lambda_e^\dagger \hfill & \phantom{space}
G^e = {18 \over 10} g^2_1+3g^2_2 \hfill \cr
N   = \lambda_\nu\lambda_\nu^\dagger \hfill & \phantom{space}
G^\nu = {6 \over 10} g^2_1+3g^2_2 \hfill } \label{UDEN}
\end{equation}
and ${\rm\ Tr}U= U_{11}+U_{22}+U_{33}$ {\it etc..} 
The general idea behind the process of decoupling the right-handed
neutrinos (in the ``step'' approximation) is that a Feynman diagram
that includes a specific flavour of a right-handed neutrino $\nu^c_A$,
with mass $M_{\nu_A}$, in an internal line only makes a
contribution to the RGEs in Eqs.~(\ref{SRNDrgeU})-(\ref{SRNDrgeN})
for energies $Q$ bigger than $M_{\nu_A}$.
Thus, the procedure depends on properly adapting
the $N$ parameter in Eq.~\refeqn{UDEN}. 
We shall now make this statement more precise. 
Let us assume that the neutrino Majorana matrix $M_{RR}$ is
diagonalized by the following transformation~:
\begin{equation}
S^{\nu^c\dagger} M_{RR} S^{\nu^c} = M'_{RR} 
= {\rm\ diag}(M_{\nu_1},M_{\nu_2},M_{\nu_3})
\end{equation}
Then, the decoupling of the right-handed neutrinos 
in Eqs.~(\ref{SRNDrgeU})-(\ref{SRNDrgeN}) can be accounted by 
replacing $N$ in Eq.~\refeqn{UDEN} by $N_\theta$ given by :
\begin{equation}
N=\lambda_\nu \lambda^\dagger_\nu =
\lambda_\nu S^{\nu^c} S^{\nu^c\dagger} \lambda^\dagger_\nu \to
\lambda_\nu S^{\nu^c} \Theta S^{\nu^c\dagger} \lambda^\dagger_\nu =  N_\theta
\end{equation}
where $\Theta(Q)$ is a energy dependent diagonal matrix defined by :
\begin{equation}
\Theta(Q) = {\rm\ diag}(
\theta(Q-M_{\nu_1}),
\theta(Q-M_{\nu_2}),
\theta(Q-M_{\nu_3}))
\end{equation}
with $\theta(x)=0$ for $x<0$ and  $\theta(x)=1$ for $x>0$.

The second issue that we would like to address concerns the effect of
a large $(\lambda_\nu)_{23}$ coupling on third family Yukawa
unification, and as a consequence, for example, on the prediction for the top mass.
We claim that the effect is small. To see why let us assume that the
only large Yukawa couplings in Eqs.~(\ref{SRNDrgeU})-(\ref{SRNDrgeN}) are 
$\lambda_t = (\lambda_u)_{33}$, $\lambda_b = (\lambda_d)_{33}$,
$\lambda_\tau = (\lambda_e)_{33}$ and
$\lambda_{\nu_\tau} = (\lambda_\nu)_{33}$,
$\lambda_{23}  = (\lambda_\nu)_{23}$. 
In this limit, the RGEs simplify to :
\begin{eqnarray}
{1 \over 16\pi^2}
{d\lambda_t \over dt} &=&
\lambda_t ( 6 \lambda_t^2 + \lambda_b^2 + \lambda_{\nu_\tau}^2 +
              \lambda_{23}^2-G^u) \label{SRNDrget} \\
{1 \over 16\pi^2}
{d\lambda_b \over dt} &=&
\lambda_b ( 6 \lambda_b^2 + \lambda_t^2 + \lambda_{\tau}^2 - G^d)
\label{SRNDrgeb} \\
{1 \over 16\pi^2}
{d\lambda_\tau \over dt} &=&
\lambda_\tau ( 4 \lambda_\tau^2 + 3 \lambda_b^2 + \lambda_{\nu_\tau}^2 - G^e ) 
\label{SRNDrgee}\\
{1 \over 16\pi^2}
{d\lambda_{\nu_\tau} \over dt} &=&
\lambda_{\nu_\tau} ( 4 \lambda_{\nu_\tau}^2 + 4 \lambda_{23}^2 +
                     3 \lambda_t^2 + \lambda_\tau^2-G^\nu) \label{SRNDrgen}
\end{eqnarray}
From Eqs.~\refeqn{SRNDrgeb},\refeqn{SRNDrgee} 
we see that the presence of the $\lambda_{23}$
coupling does not affect the RGEs of $\lambda_{b,\tau}$. 
Moreover the effect of $\lambda_{23}$ on the RGE of $\lambda_t$ 
is small ($1/8 \sim 12 \%$.)
The only RGE that is significantly affected by $\lambda_{23}$ is the
RGE of $\lambda_{\nu_\tau}$. However, since the correct prediction 
for the heaviest left-handed neutrino $m_{\nu_3}\sim 0.05 {\rm\ eV}$
requires that $M_{\nu_3} > 10^{13} {\rm\ GeV}$,
the $\lambda_{\nu_\tau}^2$ and $\lambda_{23}^2$ terms in 
Eqs.~\refeqn{SRNDrget},\refeqn{SRNDrgen} 
are only present in a rather short energy range,
{\it i.e.} between $10^{13} {\rm\ GeV} < Q < M_X \sim 10^{16} {\rm\ GeV}$. 
As a consequence the presence/absence of the neutrino Yukawa
couplings, as far as third family Yukawa unification is concerned, 
is not important.
\footnote{
Numerically we found that when $(\lambda_\nu)_{23}$ is allowed to take
values comparable with $(\lambda_\nu)_{33}$ the prediction for the top
mass roughly decreased by 1 GeV, the value of $\tan\beta$ decrease by
0.5 and the value of the unified third family Yukawa coupling at the
unification scale decreased by 0.015.}

Finally we find interesting to comment on the radiative corrections to
the neutrino atmospheric mixing angle $\theta_{23}$ between the GUT
and the $M_Z$ scale. 
It is well known \cite{BaLePa:MTanimoto} that the running of 
$\sin^2(2\theta_{23})$ can be understood from the following evolution equation~:
\begin{equation}
{1 \over 16\pi^2}
{1\over \sin^2(2\theta_{23})}
{d \sin^2(2\theta_{23}) \over dt} =
-2(\lambda_\tau^2-\lambda^2_\mu)
{ (m_{LL})^2_{33} - (m_{LL})^2_{22} \over
 [(m_{LL})_{33}-(m_{LL})_{22}]^2+4(m_{LL})^2_{23}}
\label{SRNDsin2t}
\end{equation}
which displays a resonance peak at $(m_{LL})_{33} \sim (m_{LL})_{22}$
when $(m_{LL})_{23}$ is small.
Generally, it is possible that $(m_{LL})_{33}$ starts at $Q=M_X$ bigger
than $(m_{LL})_{22}$ but, due to the third family Yukawa radiative
effects, to be driven to smaller values faster than $(m_{LL})_{22}$.
As a result, even if the initial values of $(m_{LL})_{33}$ and 
$(m_{LL})_{22}$ at $M_X$ are different, they may at some point
become comparable. If this is the case then a large $\theta_{23}$
angle can be generated radiatively from a small tree level $\theta_{23}$ at $M_X$. 
This mechanism, of amplifying $\theta_{23}$ radiatively, 
as been studied for example in Refs.~\cite{ElLeLoNa,BaLePa:MTanimoto}.
However, in these works, and as can be seen from Eq.~\refeqn{SRNDsin2t}, the
amplification is only efficient if at least the $\lambda_\tau$ Yukawa
coupling is large (about 2 or 3.) 
In our model, since we demanded top-bottom-tau Yukawa unification, the value
of the third family Yukawa coupling is rather small ($\sim 0.7$), 
thus the $\sin^2(2\theta_{23})$ is stable under radiative corrections.

\newpage

\SECTION{APPENDIX C}

In this appendix we show that is easy
to convert the results of the LMA MSW solution found in the main body of
this paper into results for the LOW solution which is also characterized by maximal 
$\nu_e\to\nu_\mu$ oscillations but smaller $\Delta m_{12}^2$ \cite{GoPe,JEllis}~:
\begin{equation}
\hskip -2cm {\rm LOW : } \quad\quad
\sin^2(2\theta_{12}) \sim 1 \qquad
\Delta m^2_{12} \sim 10^{-7} {\rm\ eV}^2
\end{equation}
The reason why we can adapt the LMA results is because,
as we showed in section III, in the SRHND approach, the $\theta_{12}$ and 
$\theta_{23}$ neutrino mixing angles come ``solely'' from the neutrino
Yukawa matrix. On the other hand, the neutrino mass spectrum depends on
the hierarchies of $\lambda_\nu$ and $M_{RR}$.
Thus, as long as we keep within the SRHND scenario,
we can change $M_{RR}$ to fit the LOW $\Delta m^2_{12}$ solution 
without that implying a significant change in $\theta_{12}$ and $\theta_{23}$.

Let as consider a LOW model with the same $U(1)_X$ flavour charges 
and the same operator matrix for $\lambda_f$ as in the LMA model, 
given by Table~\TabChargesLMA\ and  Eq.~\refeqn{lfMXLMAan},
but with a $M_{RR}$ matrix with the following structure :
\begin{equation}
{\displaystyle {M_{RR}(M_X)\phantom{_{33}} \over M_{RR}(M_X)_{33}}} 
= \left(\matrix{
A_{11} \epsilon^8 & A_{12} \epsilon^6 & A_{13} \epsilon^4 \cr
A_{21} \epsilon^6 & B_{22} \epsilon^4 & A_{23} \epsilon^2 \cr
A_{31} \epsilon^4 & A_{32} \epsilon^2 & A_{33} 
\phantom{\epsilon^2} \cr
}\right) \label{MrrMXLOWan}
\end{equation}
Comparing Eq.~\refeqn{MrrMXLOWan} with Eq.~\refeqn{MrrMXLMAan}
we see that these two equations differ only by their ``vertical'' 
$\delta$-component 
(we assumed that Eq.~\refeqn{MrrMXLOWan} has $M_{RR}^\delta\sim{\bf 1}$)
and by the numerical factor $B_{22}=1.821 \ne 1.072=A_{22}$. 
We note that the removal the $\delta$-factor in the 22 entry of 
$M_{RR}$ and the increase of the $B_{22} > A_{22}$
coefficient act to decrease $\Delta m^2_{12}$.

The Majorana matrix $M_{RR}(M_Z)$ and the neutrino
Yukawa matrix $\lambda_\nu(M_Z)$ in the LOW model resulting from 
$M_{RR}(M_X)$ in Eq.~\refeqn{MrrMXLOWan} and the Yukawa matrices
$\lambda_f(M_X)$ given by Table~\TabYukMXnumLMA\ 
(recall that we take $\lambda_f^{LOW}(M_X) = \lambda_f^{LMA}(M_X)$)
are shown in Table~\TabYukMznumLOW.
In Table~\TabMllVmnsLOW\ we present the predicted values for the 
left-handed neutrino matrix and the MNS matrix in the LOW model.
The results for the neutrino masses in the LOW model are given 
in Table~\TabNeutMassLOW.
Finally in Table~\TabAnglesLOW\ we show the values of the neutrino
mixing angles.

\vbox{
\begin{center}
\begin{tabular}{ccc}
\multicolumn{3}{c}{TABLE \TabYukMznumLOW} \cr
\noalign{\medskip}
\noalign{\hrule height\rulerheight}
\noalign{\smallskip}
\noalign{\hrule height\rulerheight}
\noalign{\medskip}
$
\lambda_\nu(M_Z)$ & $=$ & $\left(\matrix{
\phantom{-}
 6.120\times 10^{-6} &
\phantom{-}
 2.271\times 10^{-3} &
\phantom{-}
 6.368\times 10^{-2} \cr
\phantom{-}
 2.710\times 10^{-5} &
\phantom{-}
 2.736\times 10^{-3} &
\phantom{-}
 0.396 \cr
\phantom{-}
 2.925\times 10^{-5} &
-
 2.429\times 10^{-3} &
\phantom{-}
 0.565}\right)
$ \\
\noalign{\smallskip}
$
M_{RR}(M_Z)$ & $=$ & $\left(\matrix{
\phantom{-}
 2.848 \times 10^{ 9}\phantom{^1} &
\phantom{-}
 4.036 \times 10^{10} &
\phantom{-}
 7.218 \times 10^{11} \cr
\phantom{-}
 4.036 \times 10^{10} &
\phantom{-}
 2.071 \times 10^{12} &
\phantom{-}
 1.287 \times 10^{13} \cr
\phantom{-}
 7.218 \times 10^{11} &
\phantom{-}
 1.287 \times 10^{13} &
\phantom{-}
 4.551 \times 10^{14} \cr
}\right)
$ \\
\noalign{\medskip}
\noalign{\hrule height\rulerheight}
\noalign{\smallskip}
\noalign{\hrule height\rulerheight}
\end{tabular}
\end{center}

{\narrower\narrower\footnotesize\noindent
{TABLE \TabYukMznumLOW.}
Numerical values for the entries of the neutrino Yukawa
($\lambda_\nu$) and Majorana ($M_{RR}$) matrices at the $Q=M_Z$ scale in the LOW model 
($M_{RR}$ is given in GeV mass units.)
\par\bigskip}}

\vbox{
\begin{center}
\begin{tabular}{ccl}
\multicolumn{3}{c}{TABLE \TabMllVmnsLOW} \cr
\noalign{\medskip}
\noalign{\hrule height\rulerheight}
\noalign{\smallskip}
\noalign{\hrule height\rulerheight}
\noalign{\medskip}
$
{\displaystyle {m_{LL}(M_Z)\phantom{_{33}} \over m_{LL}(M_Z)_{33}}}$ & $=$ & $ 
\left(\matrix{
1.334\times 10^{-2} &
7.342\times 10^{-2} &
9.957\times 10^{-2} \cr
7.342\times 10^{-2} &
4.694\times 10^{-1} &
6.776\times 10^{-1} \cr
9.957\times 10^{-2} &
6.776\times 10^{-1} &
1 }\right) $ \\
\noalign{\smallskip}
$
V^{MNS}(M_Z)$ & $=$ & $\left(\matrix{
\phantom{-} 0.8291  & 
\phantom{-} 0.5559  &
           -0.0597 \cr
           -0.3955 &
\phantom{-} 0.6586  &
\phantom{-} 0.6402 \cr
\phantom{-} 0.3952  &
           -0.5072 &
\phantom{-} 0.7659 }\right)$ \\

\noalign{\medskip}
\noalign{\hrule height\rulerheight}
\noalign{\smallskip}
\noalign{\hrule height\rulerheight}
\end{tabular}
\end{center}

{\narrower\narrower\footnotesize\noindent
{TABLE \TabMllVmnsLOW.}
Predicted values for the left-handed neutrino mass matrix $m_{LL}(M_Z)$
in units of $m_{LL}(M_Z)_{33} = 3.567\times 10^{-3} {\rm\ eV}$ 
and for the MNS neutrino mixing matrix $V^{MNS}(M_Z)$ in the LOW model.
\par\bigskip}}

\vbox{
\begin{center}
\begin{tabular}{c}
\multicolumn{1}{c}{TABLE \TabNeutMassLOW} \cr
\noalign{\medskip}
\noalign{\hrule height\rulerheight}
\noalign{\smallskip}
\noalign{\hrule height\rulerheight}
\noalign{\medskip}
$
m_{\nu_1} = 7.29\times 10^{-9} {\rm\ eV}         \quad
m_{\nu_2} = 3.54\times 10^{-4} {\rm\ eV}         \quad
m_{\nu_3} = 5.25\times 10^{-2} {\rm\ eV}               
$ \\
\noalign{\medskip}
$
\Delta m_{12}^2 = 1.25\times 10^{-7} {\rm\ eV}^2 \qquad
\Delta m_{23}^2 = 2.76\times 10^{-3} {\rm\ eV}^2 
$ \\
\noalign{\medskip}
\noalign{\hrule height\rulerheight}
\noalign{\smallskip}
\noalign{\hrule height\rulerheight}
\end{tabular}
\end{center}

{\narrower\narrower\footnotesize\noindent
{TABLE \TabNeutMassLOW.}
Predicted values for the left-handed neutrino masses $m_{\nu_{1,2,3}}$
and squared mass splittings 
$\Delta m^2_{12} = |m_{\nu_2}^2-m_{\nu_1}^2|$,
$\Delta m^2_{23} = |m_{\nu_3}^2-m_{\nu_2}^2|$ in the LOW model.
\par\bigskip}}

\vbox{
\begin{center}
\begin{tabular}{rrr}
\multicolumn{3}{c}{TABLE \TabAnglesLOW} \cr
\noalign{\medskip}
\noalign{\hrule height\rulerheight}
\noalign{\smallskip}
\noalign{\hrule height\rulerheight}
\noalign{\medskip}
\multicolumn{3}{c}{$Q = M_X\sim 3\times 10^{16} {\rm\ GeV}$} \cr
$\sin\theta_{12} = 0.543$ &
$\sin\theta_{23} = 0.607$ &
$\sin\theta_{13} =-0.056$ \\
$\sin^2(2\theta_{12}) = 0.832$ &
$\sin^2(2\theta_{23}) = 0.931$ &
$\sin^2(2\theta_{13}) = \phantom{-}0.013$ \\
\noalign{\medskip}
\noalign{\hrule height\rulerheight}
\noalign{\medskip}
\multicolumn{3}{c}{$Q = M_{\nu_3}\sim 5\times 10^{14} {\rm\ GeV}$} \cr
$\sin\theta_{12} = 0.545$ &
$\sin\theta_{23} = 0.617$ &
$\sin\theta_{13} =-0.058$ \\
$\sin^2(2\theta_{12}) = 0.836$ &
$\sin^2(2\theta_{23}) = 0.943$ &
$\sin^2(2\theta_{13}) =\phantom{-}0.013 $ \\
\noalign{\medskip}
\noalign{\hrule height\rulerheight}
\noalign{\medskip}
\multicolumn{3}{c}{$Q = M_Z$} \cr
$\sin\theta_{12} = 0.557$ &
$\sin\theta_{23} = 0.641$ &
$\sin\theta_{13} =-0.060$ \\
$\sin^2(2\theta_{12}) = 0.856$ &
$\sin^2(2\theta_{23}) = 0.968$ &
$\sin^2(2\theta_{13}) =\phantom{-}0.014$ \\
\noalign{\medskip}
\noalign{\hrule height\rulerheight}
\noalign{\smallskip}
\noalign{\hrule height\rulerheight}
\end{tabular}
\end{center}

{\narrower\narrower\footnotesize\noindent
{TABLE \TabAnglesLOW.}
Running of the neutrino mixing angles at the unification $Q=M_X$, 
the right-handed tau neutrino mass $Q=M_{\nu_3}$ and $Z$ boson mass $Q=M_Z$
energy scales in the LOW model.
\par\bigskip}}

\newpage

\SECTION{APPENDIX D}

In this appendix we briefly present a model that explores the
possibility of a SMA MSW solution to the solar neutrino anomaly. 
Although the SMA region is disfavoured by the latest results from
the Super-Kamiokande experiment, the SMA solution is not
completely ruled out.
\footnote{Statistically, the SMA solution can still describe 
the neutrino data with a probability of 34 \%. \cite{GoPe}}
The SMA solution data indicates \cite{GoPe}:
\begin{equation}
{\rm SMA : } \qquad
\sin^2(2\theta_{12}) \sim 1.6\times 10^{-3} \qquad
\Delta m^2_{12} \sim 5\times 10^{-6} {\rm\ eV}^2
\end{equation}

In analogy with the LMA model we start by recalling that the Yukawa and
the neutrino Majorana matrices in the SMA model can be decomposed into
a ``vertical'' $\delta$-component and a ``horizontal''
$\epsilon$-component given by :
\begin{equation}
(\lambda_f)_{AB} \sim (\lambda^\delta)_{AB} (\lambda^\epsilon)_{AB}
\qquad\qquad
(M_{RR})_{AB} \sim (M_{RR}^\delta)_{AB} (M_{RR}^\epsilon)_{AB}
\label{YukMrrSMA}
\end{equation}

\vbox{
\begin{center}
\begin{tabular}{lrccccccccc}
\multicolumn{11}{c}{TABLE \TabChargesSMA} \cr
\noalign{\medskip}
\noalign{\hrule height\rulerheight}
\noalign{\smallskip}
\noalign{\hrule height\rulerheight}
\noalign{\medskip}
& &
$X_{F_1}$ & $X_{F_2}$ & $X_{F_3}$ &
$X_{F^c_1}$ & $X_{F^c_2}$ & $X_{F^c_3}$ &
$X_h$ & $X_H$ & $X_{\bar H}$ \\
\noalign{\medskip}
\noalign{\hrule height\rulerheight}
\noalign{\medskip}
$U(1)_{\bar X}$ & : &
$2$ & $0$ & $0$ &
$3$ & $2$ & $0$ &
$0$ & $0$ & $0$ \\
\noalign{\medskip}
\noalign{\hrule height\rulerheight}
\noalign{\smallskip}
\noalign{\hrule height\rulerheight}
\end{tabular}
\end{center}

{\narrower\narrower\footnotesize\noindent
{TABLE \TabChargesSMA.} 
List of the $U(1)_{\bar X}$ charges that determine the family
structure of the Yukawa and neutrino Majorana matrices in the SMA model.
\par\bigskip}}
\noindent
The $U(1)_{\bar X}$ charges of the SMA given in Table~\TabChargesSMA\
fix the ``horizontal'' structure of $\lambda^\epsilon$ and $M_{RR}^\epsilon$
in the SMA model to be :
\begin{equation}
\lambda^\epsilon
\sim
\left(\matrix{
\epsilon^5 & \epsilon^4 & \epsilon^2 \cr
\epsilon^3 & \epsilon^2 & 1 \cr
\epsilon^3 & \epsilon^2 & 1
}\right)\qquad
M_{RR}^\epsilon
\sim
\left(\matrix{
\epsilon^6 & \epsilon^5 & \epsilon^3 \cr
\epsilon^5 & \epsilon^4 & \epsilon^2 \cr
\epsilon^3 & \epsilon^2 & 1          \cr
}\right)
\label{SRNDmRLmRRepsSMA}
\end{equation}
On the other hand, the ``vertical'' structure 
of $\lambda^\delta$ and $M_{RR}^\delta$ is given by the following
operator matrices :
\begin{equation}
\lambda^\delta \sim \left(\matrix{
{\cal O}^G+{\cal O}^{\prime\prime K} &
{\cal O}^R+{\cal O}^{\prime O} &
{\cal O}^H+{\cal O}^{\prime a} \cr
{\cal O}^G+{\cal O}^{\prime\prime V} &
{\cal O}^W+{\cal O}^{\prime H} &
{\cal O}^I+{\cal O}^{\prime W} \cr
{\cal O}^M+{\cal O}^{\prime\prime V} &
{\cal O}^g+{\cal O}^{\prime T} &
1 }\right)\qquad
M_{RR}^\delta \sim {\bf 1}
\label{SRNDmRLopMtrSMA}
\end{equation}
As a result of Eqs.~\refeqn{SRNDmRLmRRepsSMA},\refeqn{SRNDmRLopMtrSMA} the
Yukawa and the neutrino Majorana matrices in Eq.~\refeqn{YukMrrSMA}
can be written as :
\begin{eqnarray}
\lambda_f(M_X) 
&=& 
\left(\matrix{
\hfil x^G_f c_{11} \delta\epsilon^5 &
\hfil x^R_f c_{12} \delta\epsilon^4 &
\hfil x^H_f c_{13} \delta\epsilon^2 \cr
\hfil x^G_f c_{21} \delta\epsilon^3 &
\hfil x^W_f c_{22} \delta\epsilon^2 &
\hfil x^I_f c_{23} \delta\phantom{\epsilon^2} \cr
\hfil x^M_f c_{31} \delta\epsilon^3 &
\hfil x^g_f c_{32} \delta\epsilon^2 &
\hfil c_{33}\phantom{\epsilon^2} \cr
}\right)+ \nonumber \\
& & \cr
& &
\left(\matrix{
0 &
x^O_f c^\prime_{12} \delta^2\epsilon^4 &
x^a_f c^\prime_{13} \delta^2\epsilon^2 \cr
0 &
x^H_f c^\prime_{22} \delta^2\epsilon^2 &
\hfil x^W_f c^\prime_{23} \delta^2\phantom{\epsilon^2} \cr
0 &
x^T_f c^\prime_{32} \delta^2\epsilon^2 &
0
}\right)+ \nonumber \\
& & \cr
& &
\left(\matrix{
x^K_f c^{\prime\prime}_{11} \delta^3\epsilon^5 & 0 & 0 \cr
x^V_f c^{\prime\prime}_{21} \delta^3\epsilon^3 & 0 & 0 \cr
x^V_f c^{\prime\prime}_{31} \delta^3\epsilon^3 & 0 & 0 
}\right) \label{lfMXSMAan} \\
& & \cr
{\displaystyle {M_{RR}(M_X)\phantom{_{33}} \over M_{RR}(M_X)_{33}}} 
&=& \left(\matrix{
C_{11} \epsilon^6 & C_{12} \epsilon^5 & C_{13} \epsilon^3 \cr
C_{21} \epsilon^5 & C_{22} \epsilon^4 & C_{23} \epsilon^2 \cr
C_{31} \epsilon^3 & C_{32} \epsilon^2 & C_{33} \phantom{\epsilon^2} \cr
}\right) \label{MrrMXSMAan}
\end{eqnarray}

In the rest of this appendix we apply the same systematic approach
used in the main part of the paper for the LMA solution 
to the SMA model.
The approximate structure of the effective matrices resulting from 
Eqs.~\refeqn{lfMXSMAan},\refeqn{MrrMXSMAan} is given in Table~\TabYukMXAprSMA.
In Table~\TabaAValuesSMA\ we give the values of the $c$,$C$ parameters
appearing in Eqs.~\refeqn{lfMXSMAan},\refeqn{MrrMXSMAan}.
In Table~\TabYukMXnumSMA\ we present the exact numerical 
values of the Yukawa and Majorana matrices at the unification scale and
in Table~\TabYukMznumSMA\ the values of the same matrices at the $M_Z$ scale.
In Table~\TabMllVmnsSMA\ we present the predicted values for the
left-handed neutrino mass and for the MNS mixing matrices.
The predictions for masses of the physical neutrinos in the SMA model
is listed in Table~\TabNeutMassSMA\ and in Table~\TabAnglesSMA\ 
we give the predictions for the neutrino mixing angles at several
energy scales. Finally, in Table~\TabmdmsSMA\, we show the predictions
for the masses of the down and strange quarks in the SMA model.

\vbox{
\vbox{
\begin{center}
\begin{tabular}{ccccc}
\multicolumn{5}{c}{TABLE \TabYukMXAprSMA} \cr
\noalign{\medskip}
\noalign{\hrule height\rulerheight}
\noalign{\smallskip}
\noalign{\hrule height\rulerheight}
\noalign{\medskip}
$
\lambda_u(M_X)$ & $\sim$ & $\left(\matrix{
\delta^3\epsilon^5 & \delta^2\epsilon^4 & \phantom{^2}\delta\epsilon^2 \cr
\delta^3\epsilon^3 & \delta^2\epsilon^2 & 0 \cr
\delta^3\epsilon^3 & \delta^2\epsilon^2 & 1 \cr}\right)$ & $\sim$ & $
\left(\matrix{
\lambda^8 & \lambda^6 & \lambda^3 \cr
\lambda^6 & \lambda^4 & 0 \cr
\lambda^6 & \lambda^4 & 1 \cr}\right)
$ \\
\noalign{\smallskip}
$
\lambda_d(M_X)$ & $\sim$ & $\left(\matrix{
\phantom{^3}\delta  \epsilon^5 &
\phantom{^2}\delta  \epsilon^4 & 
\phantom{^3}\delta  \epsilon^2 \cr
\phantom{^3}\delta  \epsilon^3 &
\phantom{^3}\delta  \epsilon^2 & 
\delta^2  \cr
\phantom{^3}\delta  \epsilon^3 & 
            \delta^2\epsilon^2 &
1 \cr}\right)$ & $\sim$ & $
\left(\matrix{
\lambda^6 & \lambda^5 & \lambda^3 \cr
\lambda^4 & \lambda^3 & \lambda^2 \cr
\lambda^4 & \lambda^4 &         1 \cr}\right)
$ \\
\noalign{\smallskip}
$
\lambda_e(M_X)$ & $\sim$ & $\left(\matrix{
\phantom{^3}\delta  \epsilon^5 &
\phantom{^3}\delta  \epsilon^4 & 
\phantom{^3}\delta  \epsilon^2 \cr
\phantom{^3}\delta  \epsilon^3 &
\phantom{^3}\delta  \epsilon^2 & 
\delta^2            \cr
\phantom{^3}\delta  \epsilon^3 &
\phantom{^3}\delta  \epsilon^2 &
1                  \cr}\right)$ & $\sim$ & $
\left(\matrix{
\lambda^6 & \lambda^5 & \lambda^3 \cr
\lambda^4 & \lambda^3 & \lambda^2 \cr
\lambda^4 & \lambda^3 &         1 \cr}\right)
$ \\
\noalign{\smallskip}
$
\lambda_\nu(M_X)$ & $\sim$ & $\left(\matrix{
\delta^3  \epsilon^5 &
\delta^2  \epsilon^4 & 
\phantom{^3}\delta\epsilon^2 \cr
\delta^3  \epsilon^3 &
\delta^2  \epsilon^2 & 
\delta               \cr
\delta^3  \epsilon^3 &
\phantom{^3}\delta    \epsilon^2 &
1                  \cr}\right)$ & $\sim$ & $
\left(\matrix{
\lambda^8 & \lambda^6 & \lambda^3 \cr
\lambda^6 & \lambda^4 & \lambda\phantom{1} \cr
\lambda^6 & \lambda^3 &         1 \cr}\right)
$ \\
\noalign{\smallskip}
$
M_{RR}(M_X)$ & $\sim$ & $\left(\matrix{
\phantom{\delta}\epsilon^6 \phantom{^3} & 
\phantom{\delta}\epsilon^5 \phantom{^3}& 
\phantom{\delta}\epsilon^3 \phantom{^3}\cr
\phantom{\delta}\epsilon^5 \phantom{^3}& 
\phantom{\delta}\epsilon^4 \phantom{^3}& 
\phantom{\delta}\epsilon^2 \phantom{^3}\cr
\phantom{\delta}\epsilon^3 \phantom{^3}& 
\phantom{\delta}\epsilon^2 \phantom{^3}& 
1\cr}\right)$ & $\sim$ & $
\left(\matrix{
\lambda^6 & \lambda^5 & \lambda^3 \cr
\lambda^5 & \lambda^4 & \lambda^2 \cr
\lambda^3 & \lambda^2 & 1\cr}\right)
$ \\
\noalign{\medskip}
\noalign{\hrule height\rulerheight}
\noalign{\smallskip}
\noalign{\hrule height\rulerheight}
\end{tabular}
\end{center}

{\narrower\narrower\footnotesize\noindent
{TABLE \TabYukMXAprSMA.}
Approximate structure of the Yukawa and neutrino Majorana matrices
in the SMA model resulting from Eqs.~\refeqn{lfMXSMAan},\refeqn{MrrMXSMAan} 
when the numerical effect of the Clebsch and of the $c$,$C$ parameters is neglected.
\par}}

\vbox{
\begin{center}
\begin{tabular}{ccc}
\multicolumn{3}{c}{TABLE \TabaAValuesSMA} \cr
\noalign{\medskip}
\noalign{\hrule height\rulerheight}
\noalign{\smallskip}
\noalign{\hrule height\rulerheight}
\noalign{\medskip}
$
c$ & $=$ & $
\left(\matrix{
-1.403 &
-1.656 &
\phantom{-}
 1.000 \cr
\phantom{-}
 1.000 &
-1.563 &
\phantom{-}
 1.000 \cr
\phantom{-}
 1.000 &
\phantom{-}
 1.000 &
\phantom{-}
 0.682  
}\right)$ \\
\noalign{\smallskip}
$
c^{\prime}$ & $=$ & $
\left(\matrix{
\phantom{-}
 0.000 &
\phantom{-}
 1.854 &
\phantom{-}
 1.000 \cr
\phantom{-}
 0.000 &
-0.807 &
\phantom{-}
 0.702 \cr
\phantom{-}
 0.000 &
\phantom{-}
 1.000 &
\phantom{-}
 0.000
}\right)$ \\
\noalign{\smallskip}
$
c^{\prime\prime}$ & $=$ & $
\left(\matrix{
-1.131 &
\phantom{-}
 0.000 &
\phantom{-}
 0.000 \cr
\phantom{-}
 1.000 &
\phantom{-}
 0.000 &
\phantom{-}
 0.000 \cr
\phantom{-}
 1.000 &
\phantom{-}
 0.000 &
\phantom{-}
 0.000
}\right)$ \\
\noalign{\smallskip}
$
C$ & $=$ & $
\left(\matrix{
\phantom{-}
1.069 &
\phantom{-}
0.533 &
\phantom{-}
0.799 \cr
\phantom{-}
0.533 &
\phantom{-}
1.054 &
\phantom{-}
0.753 \cr
\phantom{-}
0.799 &
\phantom{-}
0.753 &
\phantom{-}
1.000 
}\right)$ \\
\noalign{\medskip}
\noalign{\hrule height\rulerheight}
\noalign{\smallskip}
\noalign{\hrule height\rulerheight}
\end{tabular}
\end{center}

{\narrower\narrower\footnotesize\noindent
{TABLE \TabaAValuesSMA.}
Numerical values of the order-one $c$,$C$ parameters 
that parameterise the Yukawa and neutrino Majorana matrices of 
Eqs.~\refeqn{lfMXSMAan},\refeqn{MrrMXSMAan} in the SMA model.
\par}}}

\vbox{
\vbox{
\begin{center}
\begin{tabular}{ccc}
\multicolumn{3}{c}{TABLE \TabYukMXnumSMA} \cr
\noalign{\medskip}
\noalign{\hrule height\rulerheight}
\noalign{\smallskip}
\noalign{\hrule height\rulerheight}
\noalign{\medskip}
$
\lambda_u(M_X) $ & $=$ & $\left(\matrix{
-
9.935\times 10^{-6} &
\phantom{-}
1.880\times 10^{-4} &
\phantom{-}
1.183\times 10^{-2} \cr
\phantom{-}
1.603\times 10^{-4} &
-
1.511\times 10^{-3} &
\phantom{-}
0.000 \cr
\phantom{-}
1.603\times 10^{-4} &
\phantom{-}
1.325\times 10^{-3} &
\phantom{-}
0.682 }\right) 
$ \\
\noalign{\smallskip}
$
\lambda_d(M_X)$ & $=$ & $\left(\matrix{
-
1.423\times 10^{-4} &
-
1.365\times 10^{-3} &
\phantom{-}
7.572\times 10^{-3} \cr
\phantom{-}
2.095\times 10^{-3} &
-
1.128\times 10^{-2} &
\phantom{-}
2.149\times 10^{-2} \cr
\phantom{-}
3.313\times 10^{-3} &
\phantom{-}
 6.626\times 10^{-4} &
\phantom{-}
0.682 }\right)
$ \\
\noalign{\smallskip}
$
\lambda_e(M_X)$ & $=$ & $\left(\matrix{
-
2.846\times 10^{-4} &
-
 1.024\times 10^{-3} &
\phantom{-}
 8.518\times 10^{-3} \cr
\phantom{-}
 4.190\times 10^{-3} &
\phantom{-}
 3.006\times 10^{-2} &
-
 6.447\times 10^{-2} \cr
\phantom{-}
 3.313\times 10^{-3} &
\phantom{-}
 1.572\times 10^{-2} &
\phantom{-}
0.683}\right)
$ \\
\noalign{\smallskip}
$
\lambda_\nu(M_X)$ & $=$ & $\left(\matrix{
-
 7.451\times 10^{-6} &
\phantom{-}
 3.760\times 10^{-4} &
\phantom{-}
 1.704\times 10^{-2} \cr
\phantom{-}
 1.603\times 10^{-4} &
-
 3.023\times 10^{-3} &
\phantom{-}
 0.440 \cr
\phantom{-}
 1.603\times 10^{-4} &
\phantom{-}
 1.638\times 10^{-2} &
\phantom{-}
 0.682}\right)
$ \\
\noalign{\smallskip}
$
M_{RR}(M_X)$ & $=$ & $\left(\matrix{
\phantom{-}
 1.177 \times 10^{11} &
\phantom{-}
 2.664 \times 10^{11} &
\phantom{-}
 8.261 \times 10^{12} \cr
\phantom{-}
 2.664 \times 10^{11} &
\phantom{-}
 2.398 \times 10^{12} &
\phantom{-}
 3.541 \times 10^{13} \cr
\phantom{-}
 8.261 \times 10^{12} &
\phantom{-}
 3.541 \times 10^{13} &
\phantom{-}
 9.708 \times 10^{14} \cr
}\right)
$ \\
\noalign{\medskip}
\noalign{\hrule height\rulerheight}
\noalign{\smallskip}
\noalign{\hrule height\rulerheight}
\end{tabular}
\end{center}
\vskip -5pt
{\narrower\narrower\footnotesize\noindent
{TABLE \TabYukMXnumSMA.}
Numerical values for the entries of the Yukawa 
($\lambda_u$,$\lambda_d$,$\lambda_e$,$\lambda_\nu$)
and neutrino Majorana ($M_{RR}$) matrices at the unification scale $M_X$ in the SMA model 
($M_{RR}$ is given in GeV mass units.)
\par}}
\vskip -10pt
\vbox{
\begin{center}
\begin{tabular}{ccc}
\multicolumn{3}{c}{TABLE \TabYukMznumSMA} \cr
\noalign{\medskip}
\noalign{\hrule height\rulerheight}
\noalign{\smallskip}
\noalign{\hrule height\rulerheight}
\noalign{\medskip}
$
\lambda_u(M_Z) $ & $=$ & $ 
\left(\matrix{
-
 2.323\times 10^{-5} &
\phantom{-}
 3.837\times 10^{-4} &
\phantom{-}
 1.644\times 10^{-2} \cr
\phantom{-}
 3.409\times 10^{-4} &
-
 3.228\times 10^{-3} &
-3.588\times 10^{-3} \cr
\phantom{-}
 2.125\times 10^{-4} &
\phantom{-}
 1.767\times 10^{-3} &
\phantom{-}
 0.907 }\right)
$ \\
\noalign{\smallskip}
$
\lambda_d(M_Z) $ & $=$ & $ 
\left(\matrix{
-
 3.155\times 10^{-4} &
-
 2.734\times 10^{-3} &
\phantom{-}
 8.864\times 10^{-3} \cr
\phantom{-}
 4.130\times 10^{-3} &
-
 2.256\times 10^{-2} &
\phantom{-}
 3.088\times 10^{-2} \cr
\phantom{-}
 4.162\times 10^{-3} &
\phantom{-}
 1.053\times 10^{-3} &
\phantom{-}
 0.864 }\right)
$ \\
\noalign{\smallskip}
$
\lambda_e(M_Z) $ & $=$ & $ 
\left(\matrix{
-2.846\times 10^{-4} &
-
 1.032\times 10^{-3} &
\phantom{-}
 6.511\times 10^{-3} \cr
\phantom{-}
 4.090\times 10^{-3} &
\phantom{-}
 2.923\times 10^{-2} &
-5.322\times 10^{-2} \cr
\phantom{-}
 2.619\times 10^{-3} &
\phantom{-}
 1.257\times 10^{-2} &
\phantom{-}
 0.526 }\right)
$ \\
\noalign{\smallskip}
$
\lambda_\nu(M_Z) $ & $=$ & $ 
\left(\matrix{
-
 7.251\times 10^{-6} &
\phantom{-}
 3.290\times 10^{-4} &
\phantom{-}
 1.481\times 10^{-2} \cr
\phantom{-}
 1.464\times 10^{-4} &
-
 2.961\times 10^{-3} &
\phantom{-}
 0.399 \cr
\phantom{-}
 1.341\times 10^{-4} &
\phantom{-}
 1.387\times 10^{-2} &
\phantom{-}
 0.572 }\right)
$ \\
\noalign{\smallskip}
$
M_{RR}(M_Z)$ & $=$ & $\left(\matrix{
\phantom{-}
 1.176 \times 10^{11} &
\phantom{-}
 2.629 \times 10^{11} &
\phantom{-}
 8.034 \times 10^{12} \cr
\phantom{-}
 2.629 \times 10^{11} &
\phantom{-}
 2.370 \times 10^{12} &
\phantom{-}
 3.409 \times 10^{13} \cr
\phantom{-}
 8.034 \times 10^{12} &
\phantom{-}
 3.409 \times 10^{13} &
\phantom{-}
 9.197 \times 10^{14} \cr
}\right)
$ \\
\noalign{\medskip}
\noalign{\hrule height\rulerheight}
\noalign{\smallskip}
\noalign{\hrule height\rulerheight}
\end{tabular}
\end{center}
\vskip -5pt
{\narrower\narrower\footnotesize\noindent
{TABLE \TabYukMznumSMA.}
Numerical values for the entries of the Yukawa
($\lambda_u$,$\lambda_d$,$\lambda_e$,$\lambda_\nu$)
and neutrino Majorana ($M_{RR}$) matrices at the $Q=M_Z$ scale in the SMA model 
($M_{RR}$ is given in GeV mass units.)
\par}}}

\vbox{
\begin{center}
\begin{tabular}{ccl}
\multicolumn{3}{c}{TABLE \TabMllVmnsSMA} \cr
\noalign{\medskip}
\noalign{\hrule height\rulerheight}
\noalign{\smallskip}
\noalign{\hrule height\rulerheight}
\noalign{\medskip}
$
{\displaystyle {m_{LL}(M_Z)\phantom{_{33}} \over m_{LL}(M_Z)_{33}}}$ & $=$ & $ 
\left(\matrix{
7.530\times 10^{-4} &
2.284\times 10^{-2} &
2.741\times 10^{-2} \cr
2.284\times 10^{-2} &
8.151\times 10^{-1} &
8.239\times 10^{-1} \cr
2.741\times 10^{-2} &
8.239\times 10^{-1} &
1 }\right) $ \\
\noalign{\smallskip}
$
V^{MNS}(M_Z)$ & $=$ & $\left(\matrix{
\phantom{-} 0.9990  & 
\phantom{-} 0.0192  &
\phantom{-} 0.0407 \cr
           -0.0430 &
\phantom{-} 0.6753  &
\phantom{-} 0.7363 \cr
           -0.0133  &
           -0.7373 &
\phantom{-} 0.6754 }\right)$ \\

\noalign{\medskip}
\noalign{\hrule height\rulerheight}
\noalign{\smallskip}
\noalign{\hrule height\rulerheight}
\end{tabular}
\end{center}

{\narrower\narrower\footnotesize\noindent
{TABLE \TabMllVmnsSMA.}
Predicted values for the left-handed neutrino mass matrix $m_{LL}(M_Z)$
in units of $m_{LL}(M_Z)_{33} = 2.893\times 10^{-3} {\rm\ eV}$ 
and for the MNS neutrino mixing matrix $V^{MNS}(M_Z)$ in the SMA model.
\par\bigskip}}

\vbox{
\begin{center}
\begin{tabular}{c}
\multicolumn{1}{c}{TABLE \TabNeutMassSMA} \cr
\noalign{\medskip}
\noalign{\hrule height\rulerheight}
\noalign{\smallskip}
\noalign{\hrule height\rulerheight}
\noalign{\medskip}
$
m_{\nu_1} = 3.07\times 10^{-8} {\rm\ eV}         \quad
m_{\nu_2} = 2.27\times 10^{-3} {\rm\ eV}         \quad
m_{\nu_3} = 5.03\times 10^{-2} {\rm\ eV}               
$ \\
\noalign{\medskip}
$
\Delta m_{12}^2 = 5.15\times 10^{-6} {\rm\ eV}^2 \qquad
\Delta m_{23}^2 = 2.52\times 10^{-3} {\rm\ eV}^2 
$ \\
\noalign{\medskip}
\noalign{\hrule height\rulerheight}
\noalign{\smallskip}
\noalign{\hrule height\rulerheight}
\end{tabular}
\end{center}

{\narrower\narrower\footnotesize\noindent
{TABLE \TabNeutMassSMA.}
Predicted values for the left-handed neutrino masses $m_{\nu_{1,2,3}}$
and squared mass splittings 
$\Delta m^2_{12} = |m_{\nu_2}^2-m_{\nu_1}^2|$,
$\Delta m^2_{23} = |m_{\nu_3}^2-m_{\nu_2}^2|$ in the SMA model.
\par\bigskip}}

\vbox{
\begin{center}
\begin{tabular}{rrr}
\multicolumn{3}{c}{TABLE \TabAnglesSMA} \cr
\noalign{\medskip}
\noalign{\hrule height\rulerheight}
\noalign{\smallskip}
\noalign{\hrule height\rulerheight}
\noalign{\medskip}
\multicolumn{3}{c}{$Q = M_X\sim 3\times 10^{16} {\rm\ GeV}$} \cr
$\sin\theta_{12} = 2.17\times 10^{-2}$ &
$\sin\theta_{23} = 0.703$ &
$\sin\theta_{13} = 3.87\times 10^{-2}$ \\
$\sin^2(2\theta_{12}) = 1.87  \times 10^{-3} $ &
$\sin^2(2\theta_{23}) = 1.000                $ &
$\sin^2(2\theta_{13}) = 5.97  \times 10^{-3} $ \\
\noalign{\medskip}
\noalign{\hrule height\rulerheight}
\noalign{\medskip}
\multicolumn{3}{c}{$Q = M_{\nu_3}\sim 9\times 10^{14} {\rm\ GeV}$} \cr
$\sin\theta_{12} = 2.12\times 10^{-2}$ &
$\sin\theta_{23} = 0.713$ &
$\sin\theta_{13} = 3.94\times 10^{-2}$ \\
$\sin^2(2\theta_{12}) = 1.80  \times 10^{-3} $ &
$\sin^2(2\theta_{23}) = 1.000                $ &
$\sin^2(2\theta_{13}) = 6.19  \times 10^{-3} $ \\
\noalign{\medskip}
\noalign{\hrule height\rulerheight}
\noalign{\medskip}
\multicolumn{3}{c}{$Q = M_Z$} \cr
$\sin\theta_{12} = 1.92\times 10^{-2}$ &
$\sin\theta_{23} = 0.737$ &
$\sin\theta_{13} = 4.07\times 10^{-2}$ \\
$\sin^2(2\theta_{12}) = 1.48  \times 10^{-3} $ &
$\sin^2(2\theta_{23}) = 0.993                $ &
$\sin^2(2\theta_{13}) = 6.60  \times 10^{-3} $ \\
\noalign{\medskip}
\noalign{\hrule height\rulerheight}
\noalign{\smallskip}
\noalign{\hrule height\rulerheight}
\end{tabular}
\end{center}

{\narrower\narrower\footnotesize\noindent
{TABLE \TabAnglesSMA.}
Running of the neutrino mixing angles at the unification $Q=M_X$, 
the right-handed tau neutrino mass $Q=M_{\nu_3}$ and $Z$ boson mass $Q=M_Z$
energy scales in the SMA model.
\par\bigskip}}

\vbox{
\begin{center}
\begin{tabular}{ll}
\multicolumn{2}{c}{TABLE \TabmdmsSMA} \cr
\noalign{\medskip}
\noalign{\hrule height\rulerheight}
\noalign{\smallskip}
\noalign{\hrule height\rulerheight}
\noalign{\medskip}
$m_d(1 {\rm\ GeV}) = 7.6 {\rm\ MeV}          \quad $ &
$m_s(M_s) = 193 {\rm\ MeV}                   $ \\
\noalign{\medskip}
\noalign{\hrule height\rulerheight}
\noalign{\smallskip}
\noalign{\hrule height\rulerheight}
\end{tabular}
\end{center}

{\narrower\narrower\footnotesize\noindent
{TABLE \TabmdmsSMA.}
Predictions for the running $\overline{\rm MS}$ masses of the down ($m_d$)
and strange ($m_s$) quarks at $Q=1 {\rm\ GeV}$ and $Q=M_s$ respectively
in the SMA model ($M_s$ indicates the strange pole quark mass.)
\par\bigskip}}

\newpage


\end{document}